\documentclass{iopjournal}
\usepackage{amsmath,amssymb,amsfonts}%
\usepackage{amsthm}%
\usepackage{mathrsfs}%
\usepackage{bm}
\usepackage{url}
\usepackage{hyperref}
\usepackage{lmodern}

\begin{document}

\articletype{Tutorial} 

\title{A Quantum Dynamics Tutorial: Visualising Dynamical Decoupling Sequences on the Bloch Sphere}


\author{Charlie J. Patrickson$^{1, *}$\orcid{0000-0003-0550-6396}, Isaac Bailey$^2$, Antonio Orazzo$^2$, Luca Dellantonio, $^2$\orcid{0000-0000-0000-0000} and Isaac J. Luxmoore$^{1}$\orcid{0000-0002-2650-0842}}

\affil{$^1$Department of Engineering, University of Exeter, EX4 4QF, UK} 
\affil{$^2$Department of Physics and Astronomy, University of Exeter, EX4 4QF, UK} 

\affil{$^*$Author to whom any correspondence should be addressed.}

\email{cp728@exeter.ac.uk}


\begin{abstract} 

Dynamical decoupling protocols provide a versatile toolbox for robust control of quantum systems,  suppressing noise that would otherwise limit their performance. This tutorial aims to provide an intuitive interpretation of these protocols. Written from an experimentalist's perspective, we use the Bloch sphere picture to visualise quantum dynamics for a range of sequences. Here each dynamical decoupling operation can be viewed as an "effective field vector", whilst the phase and population of the two-level system is encoded into a "Bloch vector". The two-level dynamics are calculated by the cross product between these two vectors, such that the quantum state represented by the Bloch vector rotates around the effective field vector. Several prominent control schemes are simulated using this picture, starting with a Rabi oscillation, followed by the Ramsey sequence and a selection of pulsed and continuous dynamical decoupling techniques. These sequences are engineered to preserve control of the quantum system for as long as possible, with their efficacy often quantified by a coherence time. We clarify the definition of the longitudinal $T_1$, transverse $T_2$ and inhomogeneous $T_2^*$ coherence times, which are used across the literature. Each simulation has an accompanying animation to illustrate the protocol's two-level dynamics on the Bloch sphere. The custom simulation and Bloch sphere plotting scripts are also provided in an open access repository to allow the reader to reproduce, modify, and explore these results. Dynamical decoupling sequences have established themselves as an invaluable tool across the entire quantum technologies remit; by providing an open framework alongside this tutorial, we aim to make these protocols more accessible to new researchers discovering this thriving field.

\end{abstract}

\section{Introduction}\label{Introduction}

Quantum technologies are reaching a critical point in their development, delivering diverse innovations across sensing, computation and communications, with recent breakthroughs claiming quantum advantage in navigation \cite{Murado2025}, boson sampling \cite{Zhong2020} and secure communications \cite{Yin2017, Pittaluga2025}. There is a wide range of physical systems underlying these advances, including superconducting circuits, trapped ions, neutral atoms, photons, solid state defects, mechanical resonators, and quantum dots. Nevertheless, every architecture is susceptible to detrimental environmental interactions. The system's resilience against these interactions depends mainly on the operating temperature, coupling strength, and energy of the quantum system. In some cases, additional resilience can also be engineered by delivering fast, precise modulations to a quantum state that act before environmental perturbations take effect. This is the essential idea behind ``dynamical decoupling'' protocols. These techniques have found application across the quantum technologies remit because they are, to a large extent, agnostic of the particular environmental interaction. For example, the associated extended coherence times are routinely leveraged in quantum dots \cite{Dyte2025, Zaporski2023}, solid state spin defect \cite{PhysRevX.9.031045, Rizzato2025, PhysRevLett.132.223601}, superconducting \cite{PhysRevLett.121.220502} and trapped-ion \cite{Drmota2023} platforms. By using dynamical decoupling to align the operating frequencies of two or more qubits, entanglement operations have been realised using superconducting resonators \cite{PhysRevApplied.11.014017}, trapped ions \cite{Nnnerich2025}, solid state defects \cite{PhysRevLett.111.067601, Rovny2025}, mechanical resonators \cite{vonLpke2024}, and semiconductor quantum dots \cite{Shulman2012}. Furthermore, extended quantum lifetimes and frequency selective detection are foundational principles in colour-centre based quantum sensors \cite{Degen2017, Rizzato2025}. In essence, dynamical decoupling techniques represent an exceptionally versatile, and powerful framework for providing improved performance and additional functionality. 

This tutorial focuses on how to visualise and interpret the quantum dynamics that underpin common dynamical decoupling schemes. The article presents a series of quantum control protocols, and uses numerical simulations visualised on the Bloch sphere to interpret how these sequences interact with a generalised two-level quantum system. Imperfect control fields are simulated to illustrate the efficacy of each approach. In Sec. \ref{Two_Level_Systems} we introduce a generic two-level quantum system, defining a population and phase relative to the two constituent levels, and describe how these parameters evolve under the influence of a driving field. We introduce and formalise the Bloch sphere and explain how it can be used to visualise the dynamics of the two-level system. The framework is first used to illustrate a Rabi oscillation in Sec. \ref{Rabi_Oscillations}, where we employ a rotating reference frame to simplify the system's dynamics. In Sec. \ref{Ramsey_Protocol} we introduce the idea of using precisely timed control pulses to implement discrete  manipulations of the two-level phase and population and illustrate this with a Ramsey pulse sequence. More complex protocols, namely Hahn Echo and Carr-Purcell-Meiboom-Gill (CPMG) pulse sequences, are outlined in Sec. \ref{Pulsed_DD}. Continuous dynamical decoupling techniques, where the drive field continuously refines the phase and population of the two-level system, are an alternative to pulsed techniques. In Sec. \ref{Continuous DD} we introduce two variants, the spin lock and continuous concatenated dynamical decoupling. Finally in Sec.~\ref{Coherence_Times} we introduce and clarify the longitudinal $T_1$, transverse $T_2$ and inhomogeneous $T_2^*$ coherence times, which are widely used to benchmark the combined performance of a quantum system and the control scheme under test. All simulations and custom Bloch sphere plotting scripts are provided in an open access Github repository \cite{patrickson2026ddtutorial}, with animated Bloch sphere trajectories included in Supplementary files, and inline with an online preprint version of the article \cite{patrickson2026spinDynamicsTutorial}.

\section{The Bloch Sphere Representation of a Two-Level System}\label{Two_Level_Systems}

\begin{figure*}[h!] 
\centering
\includegraphics[width=1\columnwidth]{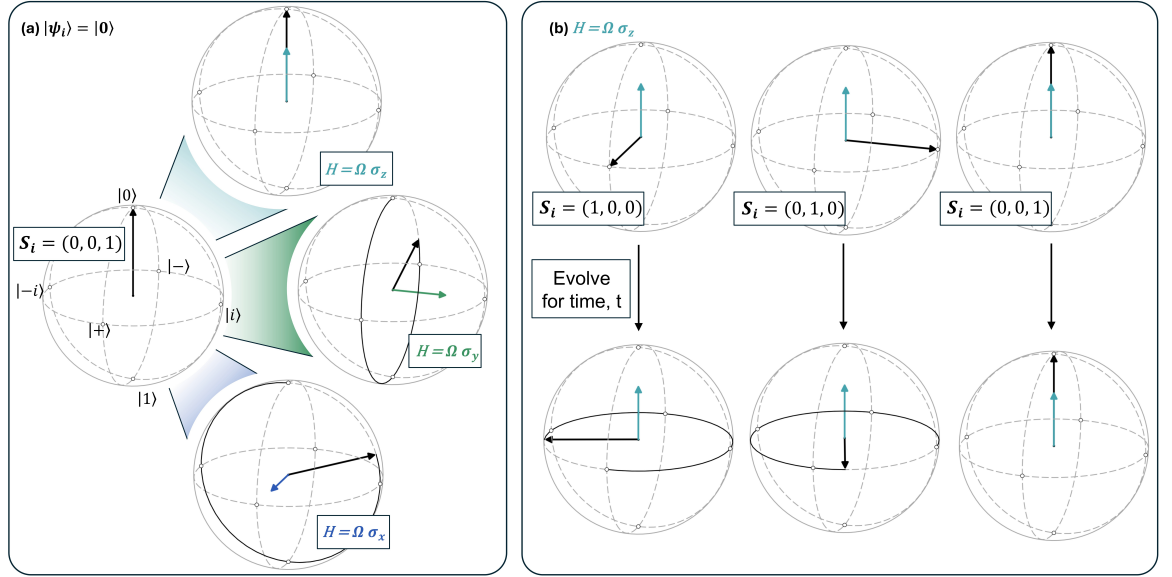}
\caption{Two-Level Dynamics on the Bloch Sphere. (a) A Bloch vector describes the state of a two-level system. The z-projection encodes the relative two-level population, where the axis intercepts represent $|0\rangle$ (+$\hat{z}$) and $|1\rangle$ (-$\hat{z}$). The relative phase is encoded in the $x$-$y$ plane, where the $\pm x$ ($\pm y$) intercepts represent the $|\pm\rangle$ ($|\pm i \rangle$) states. An effective field vector will drive the Bloch vector if it has a perpendicular component (green and blue effective field vectors). The system is in an eigenstate for a parallel effective field (turquoise effective field vector). (b) Bloch precession. Two-level systems possess an intrinsic effective field (turquoise effective field vector) that drives Bloch precession. A Bloch vector prepared in the $x$-$y$ plane (i.e. a superposition of $|0\rangle$ and $|1\rangle$) will precess around this effective field (left, middle Bloch spheres). A Bloch vector aligned with the effective field exists in an eigenstate, and does not precess (right Bloch sphere).}
\label{Fig1}
\end{figure*}

Throughout this tutorial we will use the Bloch sphere to represent the state of a two-level system. To understand it, we first define the two-level system $|\Psi\rangle$ and its constituent levels, $|0\rangle$ and $|1\rangle$. The states $|0\rangle$ and $|1\rangle$ are separated in energy by $\hbar\omega_0$. We choose $\hbar=1$ in the Hamiltonians presented later in the tutorial, corresponding to units of angular frequency rather than energy. $|\Psi\rangle$ can occupy $|0\rangle$, $|1\rangle$, or a superposition of the two, $Z_0|0\rangle + Z_1|1\rangle$. Here, $Z_{0,1}$ are complex probability amplitudes, $Z_{0,1} = |Z_{0,1}|e^{i\phi_{0,1}}$ that encode the level's phase, $\phi_{0,1}$, and magnitude, $|Z_{0,1}|$. Quantum states are defined only up to a global phase: $|\Psi\rangle$ and $e^{i\phi_0}|\Psi\rangle$ represent the same physical state for any real $\phi_0$ \cite{Nielsen_Chuang_2010}. This can be directly seen using the tools below, as $\phi_0$ leaves the Bloch sphere dynamics unchanged. For clarity, we thus use the relative phase $\phi = \phi_1 - \phi_0$ between $|0\rangle$ and $|1\rangle$ to write an arbitrary quantum state as $|\Psi\rangle = |Z_0||0\rangle + |Z_1|e^{i\phi}|1\rangle$. Additionally, as a two-level system, the probability of a measurement on $|\Psi\rangle$ returning either $|0\rangle$ or $|1\rangle$ must be unity, so that $\langle\Psi|\Psi\rangle = |Z_0|^2 + |Z_1|^2 = 1$. This equation describes a circle, such that the angle $\theta$ between $|Z_0|$ and $|Z_1|$ is the relative population of $|0\rangle$ and $|1\rangle$ at any point in time. With $\phi$ and $\theta$ being two angles, they define a unit sphere. Each point on this sphere uniquely specifies a quantum state $|\Psi\rangle$; the azimuthal angle $\phi$, encodes the relative phase between $|0\rangle$ and $|1\rangle$, and the polar angle $\theta$, encodes their relative populations. Using these definitions we can write the state of a two-level system as $|\Psi\rangle = \cos{\frac{\theta}{2}} |0\rangle + \sin{\frac{\theta}{2}}e^{i \phi} |1\rangle$, where $Z_0 = \cos{\frac{\theta}{2}}$ and $Z_1 = \sin{\frac{\theta}{2}}e^{i \phi}$. Certain values of $\theta$ and $\phi$ represent important states used throughout this tutorial, and more broadly across quantum technologies. We have already seen that $\theta=0, \pi$ describe $|0\rangle$ and $|1\rangle$, respectively. Similarly, $\theta=\pi/2$ and $\phi = 0, \pi$ describe the states $|+\rangle$ and $|-\rangle$, respectively, whilst $\theta=\pi/2$ and $\phi = \pi/2, 3\pi/2$ describe $|i\rangle$ and $|-i\rangle$, respectively (see Fig. \ref{Fig1}(a)). 

The ability to arbitrarily locate and retain any state $|\Psi\rangle$ on this sphere underpins advanced quantum protocols. To visualise this process, $|\Psi\rangle$ can be re-expressed as the Bloch vector $\bm{S}$, running from the sphere's origin to a point on the surface. We can use the above definitions of $\theta$ and $\phi$ to define $\bm{S}$ in spherical coordinates, so that $\bm{S} = \{ \sin{\theta}\cos{\phi}, \sin{\theta}\sin{\phi}, \cos{\theta} \}$. Proposed control schemes aim to manipulate $\bm{S}$ (and equivalently $|\Psi\rangle$) as a function of time and can be understood using the equation of motion (see Sec. \ref{Coherence_Times} for damping effects, bold indicates vectors)
\begin{equation}{\label{Eq_of_motion}}
    \bm{\dot{S}} =  \bm{\mathrm{\Omega_{\mathrm{eff}}}}(t)\times \bm{S}.   
\end{equation}
$\bm{\Omega_{\mathrm{\mathrm{eff}}}}(t)$ describes an effective field vector interacting with the two-level system $\bm{S}$, and represents the proposed control scheme. Instead of dealing with the Heisenberg or Schrodinger equations to interpret the time-dependency of $|\Psi\rangle$, Eq. \eqref{Eq_of_motion} allows us to use a vector cross product to interpret the time-dependency of $\bm{S}$. In other words, evolution of the two-level system $\bm{S}$ (and equiv. $|\Psi\rangle$) only depends on whether or not the effective field vector $\bm{\Omega_{\mathrm{\mathrm{eff}}}}(t)$ aligns with the initial Bloch vector $\bm{S}$. It is common to see $\bm{\Omega_{\mathrm{\mathrm{eff}}}}(t)$ presented as a Hamiltonian, $H(t)$, where the convention is to use the $\sigma_{x, y, z}$ Pauli matrices to manipulate $|\Psi\rangle$. Crucially, $H(t)$ and $\bm{\Omega_{\mathrm{\mathrm{eff}}}}(t)$ map directly onto one another, so that $H = \bm{\Omega_{\mathrm{\mathrm{eff}}}} \cdot \bm{\sigma}$, where $\bm{\sigma}$ is a vector containing the three Pauli matrices \cite{Nielsen_Chuang_2010}. This definition means that the $\hat{x}$ terms in the control scheme $\bm{\Omega_{\mathrm{\mathrm{eff}}}}(t)$ are directly equivalent to the $\sigma_x$ terms in $H(t)$. Throughout this article we will use $H$, to describe the interactions of our model two-level system, and then consider the corresponding effective field $\bm{\Omega_{\mathrm{eff}}}(t)$ to visualise dynamics on the Bloch sphere.

In Fig. \ref{Fig1}(a) we illustrate Eq. \ref{Eq_of_motion} by calculating the dynamics of different effective fields, $\bm{\Omega_{\mathrm{eff}}}(t)$ on the initial state $|\Psi_i\rangle = |0\rangle$. This corresponds to $\bm{S_i} = (0, 0, 1) = \hat{z}$ in the Bloch representation and typically represents the system immediately after the initialisation step of an experiment. We consider three effective field vectors, $\bm{\Omega_{\mathrm{eff}}} = \Omega \hat{x}$ (blue), $\bm{\Omega_{\mathrm{eff}}} = \Omega \hat{y}$ (green) and $\bm{\Omega_{\mathrm{eff}}} = \Omega \hat{z}$ (turquoise) (see also Supplementary movies 1, 2 and 3, respectively). The first two are perpendicular to the initial state $\bm{S}= \hat{z}$, so that $\bm{\dot{S}} \neq 0$ leading to a rotation around their corresponding axes and therefore a change in $\bm{S}$. When $\bm{\Omega_{\mathrm{eff}}} = \Omega \hat{z}$, $\bm{\dot{S}} = 0$ and $\bm{S}$ remains in $\bm{S_i} = \hat{z}$. We then say that $\hat{z}$ (equiv. $|0\rangle$) is an eigenstate of $\bm{\Omega_{\mathrm{eff}}}$ (equiv. $H$), as the interaction leaves $\bm{S_i}$ unchanged. 

The intrinsic energy of a two-level system is characterised by the frequency $\omega_0$. This frequency is referred to as Bloch precession or Larmor precession in spin systems. It is associated with the $\sigma_z$ Pauli operator, to define the Bloch precession Hamiltonian $H_0 = \omega_0\sigma_z$. In the Bloch representation, this corresponds to an effective field $\bm{\Omega_{\mathrm{eff}}} = \bm{\Omega_0} = \omega_0 \hat{z}$, which has eigenstates $|0\rangle$ and $|1\rangle$. In Fig. \ref{Fig1}(b) we illustrate the effect of the Bloch precession field $\bm{\Omega_0} = \omega_0 \hat{z}$ on the initial states $\bm{S_i} = (1, 0 ,0) = \hat{x}$ (left), $\bm{S_i} = (0, 1 ,0) = \hat{y}$ (middle) and $\bm{S_i} = (0, 0 ,1) = \hat{z}$ (right) (see also Supplementary Movies 4, 5 and 1, respectively). For $\bm{S_i}= \hat{x}$ and $\bm{S_i}= \hat{y}$, $\bm{\dot{S}} \neq 0$, causing the Bloch vector to rotate in the $x$-$y$ plane around the $z$-axis. However, applying $\bm{\Omega_0} = \omega_0 \hat{z}$ to $\bm{S_i} = (0, 0 ,1) = \hat{z}$ produces $\dot{\bm{S}} = 0$, describing an eigenstate. 

Crucially, all two-level systems undergo Bloch precession; any attempt to manipulate the Bloch vector must operate coherently with this ubiquitous rotation. Attempting to use a constant amplitude (DC) field for the drive, $H_{Drive} = \Omega \sigma_{x}$ (equiv. $\bm{\Omega_{Drive}} = \Omega\hat{x}$) tilts the precession axis, so that $\bm{\Omega_{\mathrm{eff}}} = \bm{\Omega_0} + \bm{\Omega_{Drive}} = \omega_0 \hat{z} + \Omega\hat{x}$. This can lead to state mixing, resulting in complex dynamics \cite{Tetienne2012}. Instead, coherent control is achieved using oscillatory fields.

\section{Rabi Oscillations}\label{Rabi_Oscillations}
A Rabi oscillation uses an oscillating (AC) field to continuously drive a transition between $|0\rangle$ and $|1\rangle$. The time it takes to drive one complete cycle from $|0\rangle \rightarrow |1\rangle \rightarrow |0\rangle$ defines the Rabi frequency, $\Omega$. This forms the bedrock of all coherent control schemes; precise measurement of the Rabi frequency informs the experimentalist how long to apply a control field for to achieve a given rotation of the Bloch vector. In most systems the Rabi frequency is linearly dependent on the amplitude of the driving field; characterising this relationship forms an essential calibration step. 

As can be seen from Eq.~\eqref{Eq_of_motion}, driving a coherent rotation of the Bloch vector $\bm{S}$ requires $|\bm{\dot{S}}| > 0$. Maintaining this condition whilst the system undergoes Bloch precession requires an AC drive, and is fulfilled by $H_{Rabi} = \Omega \cos(\omega t) \sigma_{x}$ (equiv. $\bm{\Omega_{Rabi}} = \Omega \cos(\omega t) \hat{x}$ in the Bloch representation). This is achieved by resonantly matching the drive frequency with the Bloch precession, $\omega = \omega_0$, where $\bm{\Omega_{Rabi}}$ is phase-locked with the precessing Bloch vector $\bm{S}$. This fulfills the $|\bm{\dot{S}}| > 0$ requirement, so that the effective field $\bm{\Omega_{Rabi}}$ nudges the Bloch vector across the Bloch sphere with each $\omega$ period.

\begin{figure*}[h!] 
\centering
\includegraphics[width=1\columnwidth]{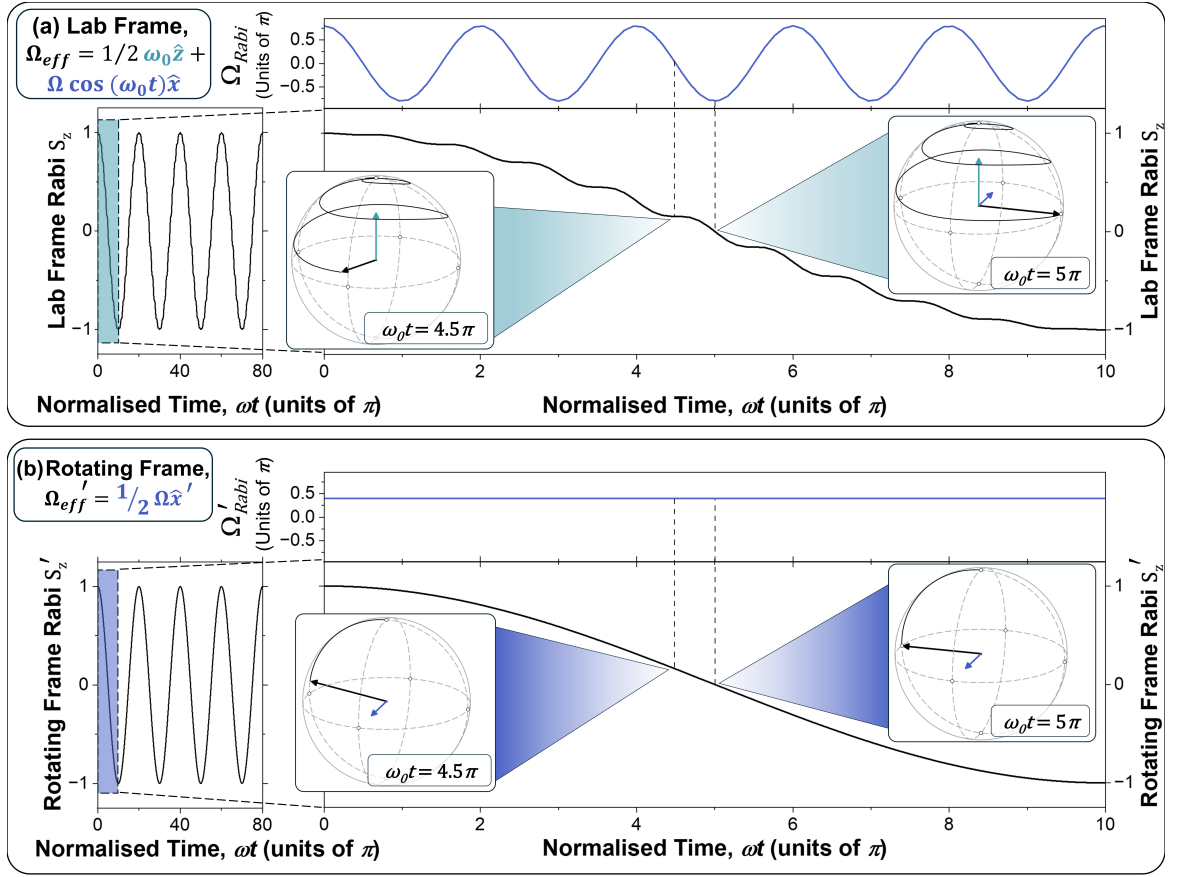}
\caption{Rabi oscillations. (a) Lab reference frame. Left, an AC field (blue) oscillating perpendicular to, and resonant with, the precessing Bloch vector (black), will drive the Bloch vector’s $S_z$ component between $|0\rangle$ ($+\hat{z}$) and $|1\rangle$ ($-\hat{z}$). The transition rate $dS_z/dt$ (zoom in, lower right) depends on the instantaneous amplitude of the applied AC field (upper right), and oscillates at the AC frequency, $\omega_0$. Bloch vector trajectories during zero and maximum instantaneous amplitudes of the AC field are illustrated on the left and right inset Bloch spheres, respectively. (b) Rotating reference frame. In a reference frame rotating around the $z$-axis at the Bloch precession frequency $\omega_0$, the AC field reduces to a constant field in $x$. The Bloch vector rotates around this DC field between +$z$ ($|0\rangle$) and -$z$ ($|1\rangle$) (left). The constant $\bm{\Omega_{\mathrm{eff}}^{\prime}}$ field amplitude (upper right) results in a constant rate of rotation around the $x^{\prime}$-axis (lower right). This is depicted on the inset Bloch spheres, which also illustrate the lack of Bloch precession in the reference frame.}
\label{Fig2}
\end{figure*}

To calculate the Bloch vector dynamics during a Rabi oscillation we use the Hamiltonian, $H = H_0 + H_{Rabi} = \tfrac{1}{2}\omega_0 \sigma_z + \Omega \cos(\omega_0t) \sigma_x$. This is described by the effective field $\bm{\Omega_{\mathrm{eff}}} = \bm{\Omega}_0 + \bm{\Omega}_{Rabi} = \tfrac{1}{2}\omega_0 \hat{z} + \Omega \cos(\omega_0t) \hat{x}$ in the Bloch representation. We choose the initial state $|\Psi_i\rangle = |0\rangle$, described by $\bm{S} = \left (0, 0, 1 \right) = \hat{z}$ on the Bloch sphere. In the left panel of Fig. \ref{Fig2}(a) we illustrate the resulting Bloch vector trajectory, plotting the $\hat{z}$ component of the Bloch vector, $S_z$, as a function of time (see also Supplementary movie 6, top panel). We see that the vector cycles between $\hat{z}$ and $-\hat{z}$, indicating continuous transitions between $|0\rangle \leftrightarrow |1\rangle$. The frequency of this oscillation defines the Rabi frequency, $\Omega$. The lower right panel provides a closeup of the turquoise highlighted region, while the upper panel plots the $\bm{\Omega}_{Rabi}$ effective field as a function of time. The closeup reveals a periodic modulation in $S_z$, which oscillates at the Rabi drive frequency $\omega_0$. This is because $\bm{\dot{S}}$ is not constant over a single Bloch precession period. For example, when $\Omega\cos(\omega_0t) \sigma_x = 0$, $\bm{\dot{S}} = 0$, as depicted on the leftmost inset Bloch sphere ($\omega_0t = 4.5\pi$) by the absence of the blue $\bm{\Omega}_{Rabi}$ effective field vector. However, when $\Omega\cos(\omega_0t) \sigma_x = \Omega \sigma_x$, $\bm{\dot{S}} = \Omega \hat{x} \times  \bm{S}$, as shown on the rightmost inset Bloch sphere ($\omega_0t = 5\pi$). In most scenarios $\omega_0 \gg \Omega$, making the modulation effect negligible. 

The Bloch sphere representation in Fig. \ref{Fig2}(a) reveals two simultaneous oscillations; a fast oscillation around the $z$-axis due to the Bloch precession term $\bm{\Omega_0} = \omega_0 \hat{z}$, and a slower oscillation around the $x$-axis due to the Rabi drive $\bm{\Omega_{Rabi}} = \Omega \cos(\omega_0t) \hat{x}$, where $\Omega<\omega_0$. The constant presence of a fast Bloch precession term can make it challenging to interpret more complex interactions. Consequently, two-level dynamics are often described using the interaction picture. In this framework, the Hamiltonian $H$ is first transformed into a rotating reference frame and then applied to the two-level system $|\Psi\rangle$\cite{10.1093/oso/9780198506348.001.0001}. In its simplest form, the reference frame is defined by an axis of rotation, and a frequency. To observe a frame that rotates with the Bloch precession, we choose the $z$-axis, and the Bloch precession frequency, $\omega_0$. In this frame, the effects of Bloch precession are no longer observed in the Bloch vector, as the rotation is implicit in the reference frame. To apply this transformation, we use the Hamiltonian, $H_0 = \tfrac{1}{2}\omega_0 \sigma_z$, to construct a transformation matrix $U = e^{iH_0t}$. This is used to calculate the rotating frame Hamiltonian, $H^{\prime} = UHU^{\dagger} + i\dot{U}U^{\dagger}$. Applying this treatment to the Rabi oscillation Hamiltonian, $H = \tfrac{1}{2}\omega_0 \sigma_z + \Omega \cos(\omega_0t) \sigma_x$, we find

\begin{align}\label{Rabi_Transformation}
    H^{\prime} &= UHU^{\dagger} + i\dot{U}U^{\dagger} \nonumber \\
    &= e^{i\tfrac{1}{2}\omega_0  t \sigma_z} \left( \tfrac{1}{2}\omega_0 \sigma_z + \Omega \cos(\omega_0  t) \sigma_x \right) e^{-i\tfrac{1}{2}\omega_0  t \sigma_z} + i \tfrac{d}{dt}e^{i\tfrac{1}{2}\omega_0  t \sigma_z} e^{-i\tfrac{1}{2}\omega_0  t \sigma_z} \nonumber \\
    &= \left(\cos(\tfrac{1}{2}\omega_0  t) + i \sin(\tfrac{1}{2}\omega_0  t)\sigma_z\right) \left( \tfrac{1}{2}\omega_0 \sigma_z + \Omega \cos(\omega_0  t) \sigma_x \right) \nonumber\\ &\phantom{=}\left(\cos(\tfrac{1}{2}\omega_0  t) - i \sin(\tfrac{1}{2}\omega_0  t)\sigma_z\right) - \tfrac{1}{2}\omega_0 \sigma_z \nonumber\\
    &=\tfrac{1}{2}\Omega \sigma_x^{\prime},
\end{align}
where primes describe operators in the rotating frame. Furthermore, we have applied the rotating wave approximation, so that fast terms oscillating at $2\omega_0$ are disregarded, and have assumed that the Rabi drive frequency matches the Bloch precession frequency, $\omega_0$. A complete derivation is included in Supplementary Note 1. Here the Bloch precession term $H_0 = \tfrac{1}{2}\omega_0 \sigma_z$ is compensated by the rotating frame, and $H_{Rabi} = \Omega \cos(\omega_0t) \sigma_x$ reduces to a constant element $H_{Rabi}^{\prime} = \tfrac{1}{2}\Omega \sigma_x$. Returning to the Bloch sphere representation, this corresponds to an effective field along $x^{\prime}$, so that $\bm{\Omega_{\mathrm{eff}}}^{\prime} = \bm{\Omega_{Rabi}}^{\prime} = \tfrac{1}{2}\Omega \hat{x}^{\prime}$. 

To calculate the Bloch dynamics in this reference frame, the initial Bloch state must also undergo the transformation. However, in this case, it remains unchanged $|\Psi_i\rangle^{\prime} = U|\Psi\rangle $$= |\Psi_i\rangle = |0\rangle$, so $\bm{S_i^{\prime}} = \left (0, 0, 1 \right) = \hat{z}^{\prime}$ on the Bloch sphere. In the left panel of Fig. \ref{Fig2}(b) we plot the $z$-component of the Bloch vector in this frame, $S_z^{\prime}$, as a function of time (see also Supplementary movie 6, bottom panel). In the absence of the Bloch precession term, $\bm{\Omega_{\mathrm{eff}}^{\prime}}$ drives a coherent rotation around the $x^{\prime}$ axis, effectively reconstructing the two-level dynamics presented above in Fig. \ref{Fig2}(a). An expanded view of the highlighted region is presented in the lower right panel, with the corresponding DC drive field $\bm{\Omega_{\mathrm{eff}}^{\prime}}$ plotted in the upper panel. Here the modulation evident in Fig. \ref{Fig2}(a) is absent. This is a consequence of removing the time dependence in $\bm{\Omega_{\mathrm{eff}}^{\prime}}$ via the rotating wave approximation. 

Whilst the Rabi oscillation is designed to continuously drive the two-level system between $|0\rangle \leftrightarrow|1\rangle$, many advanced quantum protocols use pulses to apply targeted rotations to the Bloch vector. These pulses are typically described by the angle and axis of rotation on the Bloch sphere; a rotation around the $x$-axis by an angle of $\pi$ is a $\pi_x$-pulse, a rotation around the $y$-axis by an angle of $\pi/2$ is a $\pi_y/2$-pulse (see right inset of Fig. \ref{Fig2}(b) for an example $\pi_x/2$-pulse).

\section{Ramsey Interferometry}\label{Ramsey_Protocol}

\begin{figure*}[h!] 
\centering
\includegraphics[width=1\columnwidth]{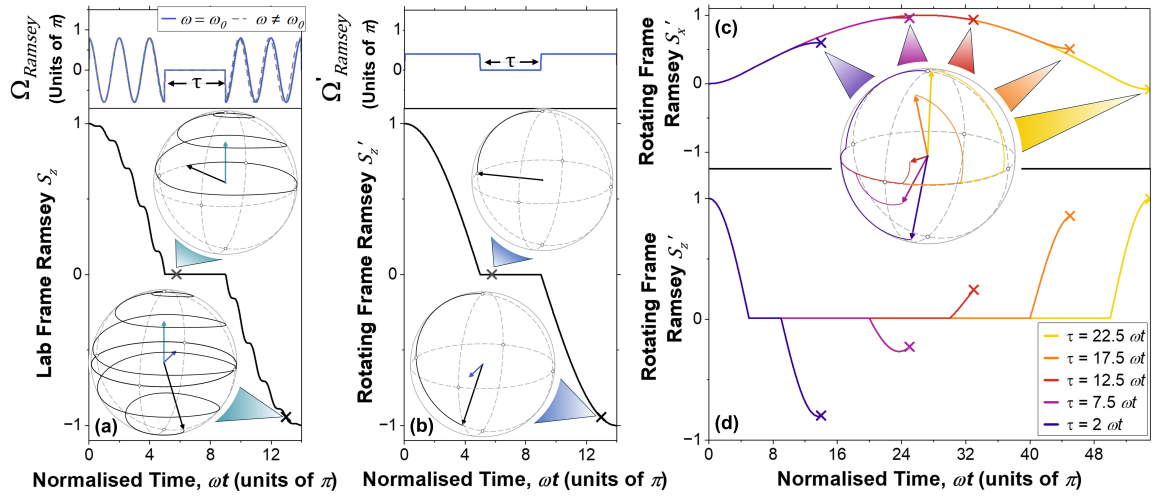}
\caption{Ramsey Interferometry. (a) Lab reference frame. Top panel, Ramsey pulse sequence. A $\tau$ delay period separates two AC $\pi_x/2$ pulses, which rotate the Bloch vector 90$^{\circ}$ about the $x$-axis. Bottom panel, Bloch vector $S_z$ trajectory. The first pulse shuttles the Bloch vector from $\hat{z}$ to the $x$-$y$ plane, where it precesses during the $\tau$ interval (top inset Bloch sphere). The second pulse projects the Bloch vector to -$\hat{z}$ (bottom inset Bloch sphere). The Bloch precession and $\pi_x/2$-pulse vectors are plotted on the Bloch spheres in turquoise and blue, respectively. (b) A reference frame rotating about the $z$-axis at the lab frame $\pi_x/2$-pulse frequency, $\omega$. Top panel, rotating frame pulse sequence. As the $\pi_x/2$-pulses are co-rotating with the reference frame, they appear as two constant fields separated by the $\tau$ delay period. Bottom panel, rotating frame Bloch vector $S_z$ trajectory. The Bloch vector appears stationary during the delay period $\tau$, as the reference frame is co-rotating with the Bloch vector at the Bloch precession frequency ($\omega = \omega_0$). (c) Applying detuned $\pi_x/2$-pulses relative to Bloch precession ($\omega \neq \omega_0$), for different $\tau$ delay periods. Top (bottom) panel, rotating frame Bloch vector $S_x^{\prime}$ ($S_z^{\prime}$) trajectories. If the frequencies of the reference frame and Bloch vector precession differ, their phases diverge, causing the Bloch vector to drift in the $x^{\prime}$-$y^{\prime}$ plane. The final $\pi_x/2$-pulse rotates the $S_x^{\prime}$ Bloch component onto $S_z^{\prime}$ before measurement, so that sweeping the $\tau$ delay period enables the accumulated Bloch phase to be sampled at multiple points (see inset Bloch sphere).}
\label{Fig3}
\end{figure*}

Ramsey interferometry uses a pulsed drive field to characterise interactions governed by the $\sigma_z$ operator, or alternatively in the Bloch sphere representation, effective fields that drive rotations around the $z$-axis. These sources perturb the natural Bloch precession frequency, and are therefore often associated with noise, or signals of interest in sensing applications. The Ramsey pulse sequence is illustrated in the upper panel of Fig. \ref{Fig3}(a). The scheme consists of an AC pulse resonant with the Bloch precession, which executes a $\pi_x/2$ rotation, followed by a delay period, $\tau$ and a second $\pi_x/2$ pulse. The initial pulse shuttles the Bloch vector from the initial state $|\Psi\rangle = |0\rangle$, corresponding to $\bm{S_i} = \left (0, 0, 1 \right) = \hat{z}$, into a superposition $|i\rangle$, or $\bm{S} = \left (0, 1, 0 \right) = \hat{y}$ on the Bloch sphere. The delay then enables a period of bare interaction with the local environment, before the second $\pi_x/2$-pulse projects the phase accumulated during $\tau$ into a measurable population difference.

We apply the sequence using the Hamiltonian $H = H_0 + H_{Ramsey} = \tfrac{1}{2}\omega_0 \sigma_z + \Omega(t) \cos(\omega t) \sigma_x$, corresponding to an effective field of $\bm{\Omega_{\mathrm{eff}}} = \bm{\Omega_0} + \bm{\Omega_{Ramsey}} = \tfrac{1}{2}\omega_0 \hat{z} + \Omega(t) \cos(\omega t) \hat{x}$. The resulting $z$-component of the Bloch vector trajectory, $S_z$, is presented in the lower panel of Fig. \ref{Fig3}(a) (see also Supplementary movie 7, upper panel). During the $\pi_x/2$ pulses we set $\Omega(t) = \Omega$ for a pulsewidth of $T = \frac{\pi}{2\Omega}$, and set $\Omega(t) = 0$ during the delay, $\tau$. We assume a resonant drive, where $\omega = \omega_0$. After the initial $\pi_x/2$-pulse, $S_z = 0$, placing the Bloch vector in the $x$-$y$ plane. The top inset Bloch sphere illustrates the trajectory, revealing the Bloch precession produced by $\bm{\Omega_0} = \omega_0 \hat{z}$, also evident from the modulation in $S_z$. In the delay period $\tau$, $\bm{\Omega_{Ramsey}} = 0$ and the Bloch vector remains in the $x$-$y$ plane precessing under $\bm{\Omega_0} = \omega_0 \hat{z}$. This is evidenced on the bottom inset Bloch sphere, which also plots the trajectory under the final $\pi_x/2$-pulse. Note that typically the experimental waveform used for the second $\pi_x/2$-pulse is a continuation of the first $\pi_x/2$-pulse. In other words, the two pulses remain phase coherent across the delay period $\tau$. 

In Fig. \ref{Fig3}(b) we repeat the analysis in a rotating reference frame that tracks with the Bloch precession, $\bm{\Omega_0} = \tfrac{1}{2}\omega_0 \hat{z}$ (see also Supplementary movie 7, lower panel). The rotating frame Hamiltonian is $H^{\prime} = \tfrac{1}{2}\Omega(t) \sigma_x^{\prime}$ (equiv. $\bm{\Omega_{\mathrm{eff}}^{\prime}} = \bm{\Omega_{Ramsey}^{\prime}}=  \tfrac{1}{2}\Omega(t) \hat{x}^{\prime}$) whilst the pulse and delay timings remain unchanged from the lab frame. The upper panel plots the $x^{\prime}$ component of the rotating frame drive field $\bm{\Omega_{Ramsey}^{\prime}}$ as a function of time, whilst the lower panel plots the rotating frame $z^{\prime}$ component of the Bloch vector, $S_z^{\prime}$. Under a resonant drive field, where $\omega = \omega_0$, the Bloch vector remains static during the delay period $\tau$. This is evidenced on the inset Bloch spheres, and is a consequence of the reference frame being phase locked with the Bloch precession.

If $\omega \neq \omega_0$ the Bloch vector dynamics are altered and depend on the frequency mismatch, or detuning, $\Delta = \omega - \omega_0$. This can be used to measure the Bloch precession frequency, $\omega_0$, as the drive frequency $\omega$ is typically a robust experimental quantity precisely controlled by the observer. To demonstrate this effect we introduce a detuned $\pi_x/2$-pulse in Fig. \ref{Fig3}(a), plotted in dashed grey. The pulse slowly loses phase coherence with the $\omega = \omega_0$ resonant blue drive. Notably, when the detuned drive switches back on for the second pulse, it is no longer perpendicular to the Bloch vector, and therefore does not drive a complete $\pi_x/2$-pulse (see Fig. \ref{Fig3}(c), purple). To interpret these dynamics we consider a reference frame defined with respect to the drive frequency $\omega$, as it represents a well-defined experimental reference. This definition introduces an additional term, $H_{\Delta}^{\prime} = \tfrac{1}{2}(\omega_0 - \omega)\sigma_z^{\prime} = \tfrac{1}{2}\Delta\sigma_z^{\prime}$, to the rotating frame Hamiltonian, $H^{\prime}$ (see Supplementary Note 1 for the full derivation). The corresponding effective field is $\bm{\Omega_{\Delta}^{\prime}} = (\omega_0 - \omega)\hat{z}^{\prime}= \tfrac{1}{2}\Delta\hat{z}^{\prime}$, so that a nonzero $\bm{\Omega_{\Delta}^{\prime}}$ term produces an additional rotation around the $z^{\prime}$-axis, causing the reference frame to lose synchronisation with the Bloch precession.

To measure the detuning between the Bloch precession and the drive frequency,$\Delta$, we sweep the Ramsey delay time, $\tau$\cite{Debroux2021, Stern2024}. The sequence is simulated with the rotating frame Hamiltonian, $H^{\prime} = H_{\Delta}^{\prime} + H_{Ramsey}^{\prime} = \tfrac{1}{2}(\omega_0 - \omega)\sigma_z^{\prime} + \tfrac{1}{2}\Omega(t) \sigma_x^{\prime}$ for a range of $\tau$ delay periods. The $S_x^{\prime}$ component of the Bloch vector is presented in the upper panel of Fig. \ref{Fig3}(c) (see also Supplementary movie 8). The initial $\pi_x/2$-pulse rotates the Bloch vector from $|0\rangle \rightarrow |i\rangle$, or $\bm{S_i^{\prime}} = \hat{z}^{\prime} \rightarrow \bm{S^{\prime}} = \hat{y}^{\prime}$ on the Bloch sphere. During the $\tau$ delay period the Bloch vector rotates around the $z^{\prime}$-axis due to nonzero detuning, $\bm{\Omega_{\Delta}^{\prime}} = \Delta\hat{z}^{\prime} \neq0$, shown on the inset Bloch sphere. This manifests as an oscillation in the $S_x^{\prime}$ component at the detuning frequency $\Delta$, representing an accumulation of phase difference between the Bloch vector and reference frame as $\tau$ is increased. The final $\pi_x/2$-pulse projects the $S_x^{\prime}$ component onto $S_z^{\prime}$, enabling the accumulated phase to be measured experimentally via the population. This is illustrated by the $S_z^{\prime}$ trajectory in the lower panel. 

Where the drive frequency $\omega$ represents a stable quantity, the Bloch precession frequency may fluctuate if the two-level system is exposed to noise. Potential sources include local magnetic fields in spin systems, such as other nearby spins, or fluctuating electric fields from laser noise in optical systems. This results in nonzero detuning, leading to uncontrolled rotations of the Bloch vector around the $z$-axis relative to the drive frequency. Notably, these noise sources can fluctuate on timescales shorter than a single measurement, so that each repeated measurement accumulates a different rotation, and therefore projects onto a different final $S_z$ value. Averaged over many measurements, this manifests as either a broadening of the probed precession frequency, or, equivalently, a decaying signal as the measurement duration increases. (see Sec. \ref{Coherence_Times}). 

\section{Pulsed Dynamical Decoupling Sequences}\label{Pulsed_DD}

Uncontrolled rotations of the Bloch vector, like those produced by a noisy $\bm{\Omega_{\Delta}^{\prime}}$, can dramatically alter the intended outcome of a control sequence. Many schemes are designed to mitigate these effects, and fall under the description of dynamical decoupling techniques. This name reflects carefully optimised, dynamic rotations of the Bloch vector that are designed to decouple two-level systems from noise. In the following we provide an overview of some of the most common approaches. Note that the Bloch vector dynamics are all calculated in a reference frame that rotates around the $z$-axis, tracking the drive field frequency, $\omega$.

\subsection{Hahn Echo}\label{Hahn_Echo}

\begin{figure*}[h!] 
\centering
\includegraphics[width=1\columnwidth]{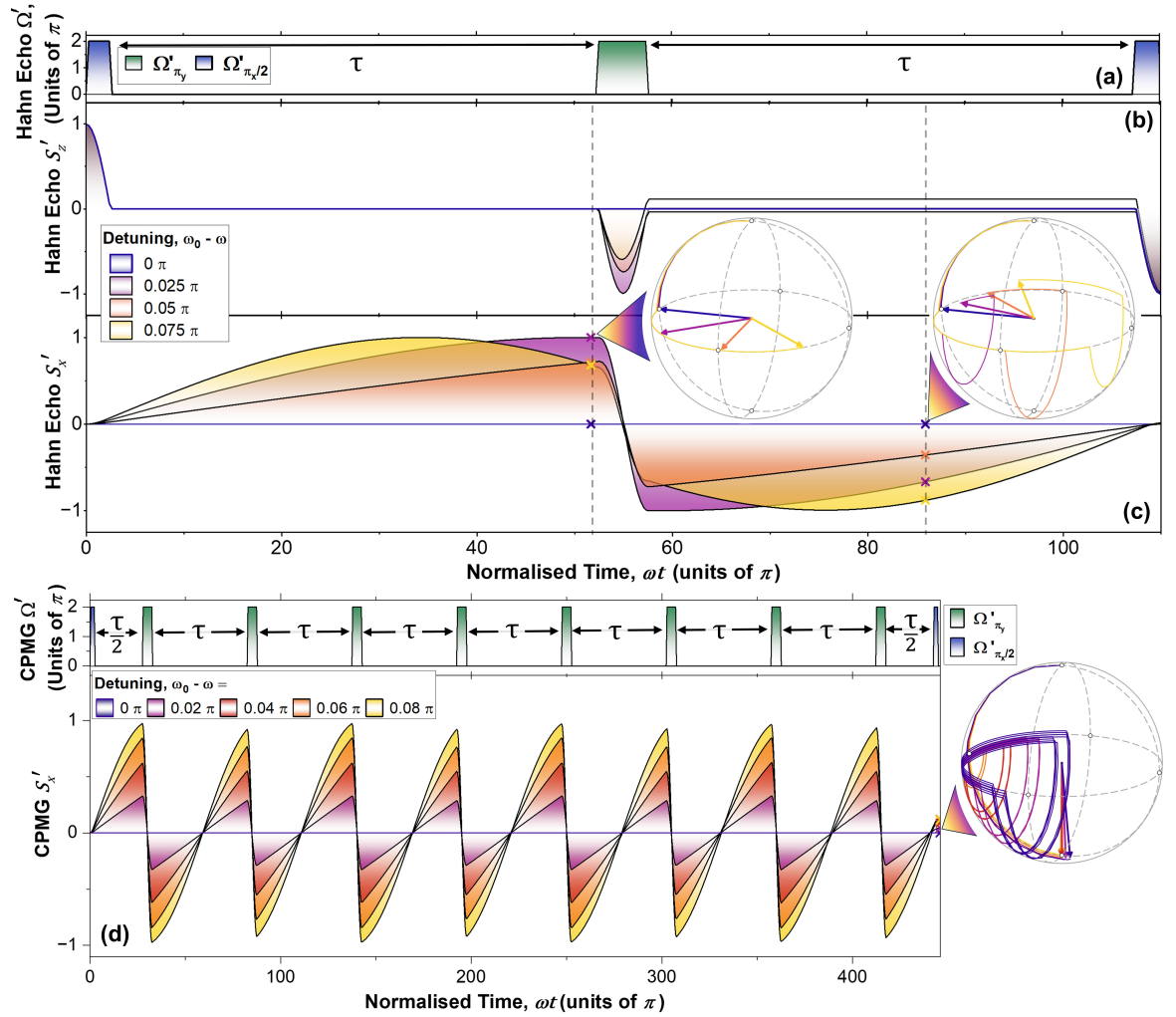}
\caption{Pulsed Dynamical Decoupling in a rotating reference frame. (a) Hahn Echo pulse sequence. Pulsed $\Omega_{\pi_x/2}^{\prime}$ ($\Omega_{\pi_y}^{\prime}$) control fields rotate the Bloch vector by 90$^{\circ}$s ($180^{\circ}$s) around the $x$-axis ($y$-axis). Each pulse is separated by a $\tau$ delay period. Due to the co-rotation of the reference frame and AC fields, $\Omega_{\pi_x/2}^{\prime}$ and $\Omega_{\pi_y}^{\prime}$ manifest as DC fields. (b, c) $S_z^{\prime}$ (b) and $S_x^{\prime}$ (c) Bloch vector trajectories, shown for a range of $\omega_0 - \omega$ detunings. After the initial $\pi_x/2$-pulse, if the reference frame ($\omega$) and Bloch vector precession frequencies ($\omega_0$) differ, the phases of each diverge. Shown here for a range of detuned $(\omega_0 - \omega)\hat{z}$ effective fields that rotate each Bloch vector in the $x$-$y$ plane at a different rate (left inset Bloch sphere). After a central $\pi_y$-pulse inverts $S_x^{\prime}$ the phases reconverge, and the detuned Bloch vectors refocus before the final $\pi_x/2$-pulse (right inset Bloch sphere). (d) Upper panel, CPMG pulse sequence. Initial and final $\pi_x/2$-pulses are applied via the DC $\Omega_{\pi_x/2}^{\prime}$ effective field. This is interceded by a train of $\pi_y$-pulses via the DC $\Omega_{\pi_y}^{\prime}$ effective field, with each separated by a $\tau$ delay period. Lower panel, $S_x^{\prime}$ CPMG trajectories for a range of $\omega_0 - \omega$ detunings. The $S_x^{\prime}$ CPMG dynamics extend the Hahn echo sequence, where each $\pi_y$ pulse refocuses the diverging Bloch vectors. This repeated sequence is shown on the inset Bloch sphere, illustrating the protocols extended ability to reverse dephasing under a constant $\omega_0 - \omega$ detuning offset.}
\label{Fig4}
\end{figure*}

The Hahn Echo, illustrated in the upper panel of Fig. \ref{Fig4}(a), is an archetypal dynamical decoupling sequence~\cite{Hahn1950} and is commonly used thanks to its relative simplicity, and its ability to characterise noise sources~\cite{Rovny2025, Shofer2025, Rizzato2025}. Like the Ramsey sequence, it begins and ends with $\pi_x/2$-pulses enclosing a free-evolution period, but inserts a refocusing $\pi_y$-pulse at the midpoint, dividing the total evolution time into two equal intervals of duration $\tau$. The $\pi_y$-pulse reverses the sign of any phase accumulated during the first interval. This has the intended effect of canceling any subsequent phase accumulation in the second interval. Notably this cancellation only occurs for phase generated by quasi-static or slowly varying fields, which remain approximately constant over the total time $2\tau$. The Hahn echo therefore suppresses the low-frequency dephasing to which Ramsey interferometry is maximally sensitive, whilst retaining sensitivity to fields that fluctuate on timescales comparable to $\tau$. 

To analyse the Hahn Echo, we use the Hamiltonian, 

\begin{equation}\label{Hahn_Echo}
H^{\prime} = H_{\Delta}^{\prime} + H_{HE}^{\prime} = (\omega_0 - \omega)\sigma_z^{\prime} + \tfrac{1}{2}\Omega_{\pi_x/2}(t) \sigma_x^{\prime}+ \tfrac{1}{2}\Omega_{\pi_y}(t) \sigma_y^{\prime},
\end{equation}
corresponding to an effective field,
\begin{equation}
\bm{\Omega^{\prime}} = \bm{\Omega_{\Delta}^{\prime}} + \bm{\Omega_{HE}^{\prime}} = (\omega_0 - \omega)\hat{z}^{\prime} + \tfrac{1}{2}\Omega_{\pi_x/2}(t) \hat{x}^{\prime}+ \tfrac{1}{2}\Omega_{\pi_y}(t) \hat{y}^{\prime},
\end{equation}
and choose a $\pi_x/2$-pulse ($\pi_y$-pulse) pulsewidth of $T = \frac{\pi}{2\Omega_{\pi_x/2}}$ ($T = \frac{\pi}{\Omega_{\pi_y}}$), during which $\Omega_{\pi_x/2}(t) = \Omega$ ($\Omega_{\pi_y}(t) = \Omega$), and set $\Omega_{\pi_x/2}(t), \Omega_{\pi_y}(t) = 0$ during the $\tau$ delay periods. We apply the sequence for a range of ${H_{\Delta}^{\prime}}$ values, emulating a distribution of Bloch precession frequencies characteristic of an inhomogeneously broadened ensemble of systems (see Sec. \ref{T_2*}).

The Hahn echo $S_{z}^{\prime}$ and $S_{x}^{\prime}$ Bloch vector components are plotted as a function of time in Fig. \ref{Fig4}(b) and Fig. \ref{Fig4}(c), respectively (see Supplementary movie 9 also). The initial $\pi_x/2$-pulse rotates the Bloch vectors from an initial state $|\Psi_i\rangle = |0\rangle$, or $\bm{S_i} = \hat{z}$, to $|\Psi\rangle = |i\rangle$, or $\bm{S} = \hat{y}$ on the Bloch sphere. During the first $\tau$ delay period $\bm{\dot{S}_{z,\ HE}} = 0$, whilst $\bm{\dot{S}_{x,\ HE}} \neq 0$ and the $\bm{S_{x,\ HE}^{\prime}}$ components with different detunings diverge across the $x$-$y$ plane. This is illustrated on the left inset Bloch sphere, where we see that nonzero $\bm{\Omega_{\Delta}^{\prime}}$ terms cause the Bloch vectors to fan out. The central $\pi_y$-pulse simultaneously inverts these $S_{x,\ HE}^{\prime}$ components, causing them to re-converge during the second $\tau$ delay period (see right inset Bloch sphere). Unlike the Ramsey sequence, this engineered refocusing of the Bloch vector effectively compensates the range of uncontrolled rotations. The final $\pi_x/2$ pulse projects the converged Bloch vectors onto the $z$-axis, where they complete the sequence in the target state $|\Psi_f\rangle^{\prime} = |1\rangle$, corresponding to $\bm{S_f}^{\prime} = (0, 0, -1) = -\hat{z}^{\prime}$ in the Bloch representation. Note that for a resonant Hahn echo sequence where $\bm{\Omega_{\Delta}^{\prime}} = 0$ (plotted in dark blue) $\bm{\dot{S}} = 0$ during the central $\pi_y$-pulse. This is because after the initial $\pi_x/2$-pulse the Bloch vector lies on the $y$-axis parallel to the $\pi_y$-pulse, describing an eigenstate of the corresponding $\pi_y$ operation.

\subsection{Carr-Purcell-Meiboom-Gill (CPMG)}\label{CPMG}

It is desirable to prolong coherent control over the Bloch vector for as long as possible. Based on our noise model so far, this might appear achievable over extended timescales simply by increasing the $\tau$ delay period of the Hahn echo. That model treated the environmental interaction as static, producing a constant detuning $H^{\prime}_{\Delta}$ (equiv. $\bm{\Omega^{\prime}_{\Delta}}$) that drives the Bloch vector around the z-axis at a fixed rate. Under this assumption, the central $\pi_y$-pulse inverts the sign of the accumulated phase, so that the Bloch vector returns to its initial position regardless of $\tau$. However, the environment is rarely static. Stochastic interactions cause the precession frequency $\omega_0$ to fluctuate around its central value, so that $H^{\prime}_{\Delta}$ drifts over the course of a measurement. If this drift is appreciable over $\tau$, the phase accumulated before the $\pi_y$-pulse is no longer equal and opposite to that accumulated after it, and refocusing is incomplete. Extending $\tau$ therefore increases exposure to these fluctuations, causing a progressive loss of coherence (see Sec. \ref{T_2}). The CPMG sequence offers a solution to this problem, by replacing the single Hahn echo $\pi_y$-pulse with a train of $N$ equally spaced $\pi_y$-pulses. The total sequence length can be extended by adding additional $\pi_y$-pulses without needing to increase interpulse $\tau$ delay lengths. This means each sub-interval can minimise the timescale available for the Bloch vector to drift, enabling the following $\pi_y$-pulse to effectively refocus any phase accumulated. This allows the total interrogation time to be extended well beyond typical Hahn echo timescales, making it (and related pulse sequences) a foundational tool in quantum control and sensing.

A CPMG sequence is illustrated in the upper panel of Fig. \ref{Fig4}(d), starting and finishing with two $\pi_x/2$-pulses (blue), with eight $\pi_y$-pulses (green) in between. Note that the first and last delay periods are halved to $\tau/2$ in the CPMG protocol, making the total evolution time $N\tau$. Our simulation uses the Hamiltonian $H^{\prime} = H_{\Delta}^{\prime} + H_{CPMG}^{\prime} = (\omega_0 - \omega)\sigma_z^{\prime} + \tfrac{1}{2}\Omega_{\pi_x/2}(t) \sigma_x^{\prime}+ \tfrac{1}{2}\Omega_{\pi_y}(t) \sigma_y^{\prime}$, described by the effective field $\bm{\Omega^{\prime}} = \bm{\Omega_{\Delta}^{\prime}} + \bm{\Omega_{CPMG}^{\prime}} = (\omega_0 - \omega)\hat{z}^{\prime} + \tfrac{1}{2}\Omega_{\pi_x/2}(t) \hat{x}^{\prime}+ \tfrac{1}{2}\Omega_{\pi_y}(t) \hat{y}^{\prime}$, with $\pi_x/2$-pulse ($\pi_y$-pulse) pulsewidths of $T = \frac{\pi}{2\Omega_{\pi_x/2}}$ ($T = \frac{\pi}{\Omega_{\pi_y}}$), during which $\Omega_{\pi_x/2}(t) = \Omega$ ($\Omega_{\pi_y}(t) = \Omega$). We again apply the sequence for a range of ${H_{\Delta}^{\prime}}$ values, and fix the $\tau$ delay period during which $\Omega_{\pi_x/2}(t), \Omega_{\pi_y}(t) = 0$. 

In the lower panel of Fig. \ref{Fig4}d) we plot the $S_x^{\prime}$ components of the CPMG sequence as a function of time (see Supplementary movie 10 also). Prior to each $\pi_y$-pulse the $S_x^{\prime}$ components scatter, describing a loss of phase coherence between the reference frame and Bloch precession. This is caused by the range of $(\omega_0 - \omega)\hat{z}^{\prime}$ detuning terms leading to different rotation speeds around the $z^{\prime}$-axis, causing the Bloch vectors to fan out in the $x^{\prime}$-$y^{\prime}$ plane. After each $\pi_y$-pulse, the Bloch vectors refocus towards the $y^{\prime}$-axis where $S_x^{\prime} = 0$, before scattering again prior to the next $\pi_y$-pulse, when the sequence repeats. This pattern is illustrated on the right inset Bloch sphere. Each $\pi_y$-pulse flips the scattering Bloch vectors to the far side of the Bloch sphere, where they reconverge on the $y^{\prime}$-axis and the sequence repeats. Note that the detuned drive field $\bm{\Omega_{CPMG}^{\prime}}$ results in a slightly different trajectory following each $\pi_y$-pulses. Nevertheless, experimental demonstrations can often reach 100's of $\pi_y$-pulses \cite{Rizzato2025}, illustrating the robustness of the CPMG sequence.  

\section{Continuous Dynamical Decoupling Sequences}\label{Continuous DD}

A wide range of pulsed dynamical decoupling techniques have been developed \cite{Degen2017, Choi2020}; however their ability to extend the coherence time of a two-level system is ultimately limited by the intrinsic lifetimes of the system (see Sec. \ref{Coherence_Times}), and long sequences place stringent demands on pulse area calibration \cite{Biercuk2009}. Continuous control schemes offer an alternative, where an uninterrupted control field is applied to continuously suppress dephasing of the Bloch vector. In the following section we introduce two variants, the spin-lock and continuous concatenated dynamical decoupling (CCDD). 

\subsection{The Spin-Lock Sequence}\label{Spin-Lock}

The spin-lock sequence applies a control field parallel to the Bloch vector and resonant with the Bloch precession \cite{10.1093/oso/9780198506348.001.0001}. In this configuration, the two-level system becomes ``locked'' to the control field, forcing the system to remain in an eigenstate of the drive. The spin lock can protect systems for a very long time, making the technique useful for slow operations such as the detection of very weak fields in quantum sensing \cite{Rizzato2025}, entangling gates between weakly coupled systems \cite{PhysRevLett.111.067601}, or as an extended idling state \cite{Bodey2019}. To execute the scheme, a $\pi_x/2$-pulse first rotates the Bloch vector from the initial state $\bm{S_i} = \left (0, 0, 1 \right) = \hat{z}$ to $\bm{S} = \left (0, 1, 0 \right) = \hat{y}$ on the Bloch sphere. The resonant drive field is immediately turned on, and applied parallel to the Bloch vector along $y$. A final $\pi_x/2$-pulse projects the Bloch vector back onto the $z$-axis for readout. 

\begin{figure*}[h!] 
\centering
\includegraphics[width=1\columnwidth]{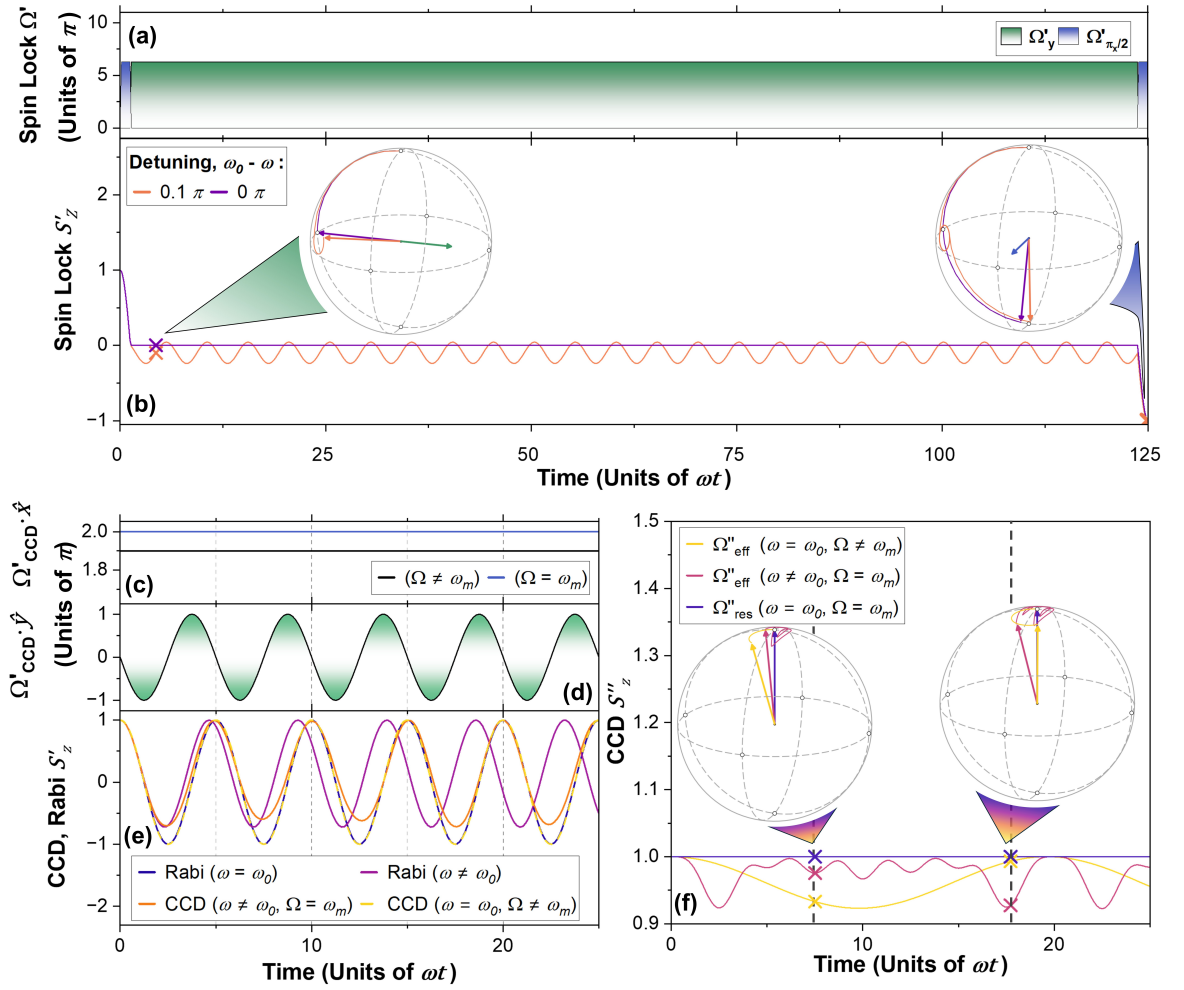}
\caption{Continuous Dynamical Decoupling. (a) Spin-lock pulse sequence in the rotating frame. $\pi_x/2$-pulses are applied at the start and end of the sequence (blue), with a continuous $\Omega_{y}^{\prime}$ field applied in between (green). (b) Rotating frame $S_z^{\prime}$ spin-lock trajectory. After the initial $\pi_x/2$-pulse transfers the Bloch vector from $\bm{S^{\prime}_i} = \hat{z}^{\prime}$ to $\bm{S^{\prime}} = \hat{y}^{\prime}$, the $\Omega^{\prime}_{y}$ field (green, left inset Bloch sphere) locks the Bloch vector to the $y^{\prime}$-axis (purple, inset Bloch spheres). If the drive field is detuned from the Bloch precession, $(\omega \neq \omega_0)$, the locking axis is instead tilted in $z^{\prime}$ (left inset Bloch sphere, orange vector). The locked Bloch vectors are projected onto $\bm{S^{\prime}} = \hat{z^{\prime}}$ by the final $\pi_x/2$-pulse (blue, right inset Bloch sphere). (c), (d) Rotating frame CCDD $x^{\prime}$ (c) and $y^{\prime}$ (d) drive components. This first reference frame rotates at the drive frequency $\omega$ around the $z$-axis. The $x^{\prime}$-components are shown for resonant ($\Omega = \omega_m$, blue) and detuned ($\Omega \neq \omega_m$, black) CCDD drive fields, with identical $y^{\prime}$-components. (e) Rotating frame $S_z^{\prime}$ trajectories for different Rabi and CCDD drive conditions. We define the target Bloch vector trajectory using a Rabi oscillation where the drive is resonant with the Bloch precession ($\omega = \omega_0$) (blue). If left uncontrolled, a detuned Rabi oscillation ($\omega \neq \omega_0$) will diverge from this trajectory (purple). The CCDD scheme compensates against both Bloch precession ($\omega \neq \omega_0$, $\omega_m = \Omega$) and Rabi frequency detunings ($\omega = \omega_0$, $\omega_m \neq \Omega$), locking the Bloch vector to the target Rabi frequency (orange and yellow, respectively). (f) $S_z^{\prime\prime}$ CCDD trajectories in a doubly rotating reference frame. This reference frame also tracks the $\omega_m$ Rabi oscillation around the $x^{\prime}$ axis. A perfectly resonant CCDD drive ($\omega = \omega_0$, $\omega_m = \Omega$) locks the Bloch vector to $\hat{z^{\prime \prime}}$ (blue, inset Bloch spheres). If the modulation is detuned from the Rabi frequency ($\omega = \omega_0$, $\omega_m \neq \Omega$), this locking axis is tilted in $x^{\prime \prime}$ (yellow, inset Bloch spheres). Detuned control and Bloch precession frequencies ($\omega \neq \omega_0$, $\omega_m = \Omega$) results in a small oscillation about  $\tfrac{1}{2}\epsilon_m \hat{z}^{\prime \prime}$ (red, inset Bloch spheres).}
\label{Fig5}
\end{figure*}

We illustrate the scheme in the same reference frame that tracks the Bloch precession, rotating around the $z$-axis at the drive frequency $\omega$. Starting from the lab frame Hamiltonian, we find the Bloch precession term, $H_0$, the $\pi_x/2$ pulse $H_{\pi_x/2}(t)$, and the spin lock field, $H_{Drive}(t)$,
\begin{align*}
H_{SpinLock} = H_{0} + H_{\pi_x/2}(t) + H_{y}(t)= \tfrac{1}{2}\omega_0\sigma_z + \Omega_{\pi/2} (t) \cos (\omega t) \sigma_x + \Omega_{y}(t) \cos (\omega t + \pi/2) \sigma_x.
\end{align*}
After switching to the reference frame defined by $U = e^{i\tfrac{1}{2}\omega_0  t \sigma_z}$,
\begin{align*}
H_{SpinLock}^{\prime} = H_{\Delta}^{\prime} + H_{\pi_x/2}^{\prime}(t) + H_{y}^{\prime}(t)= \tfrac{1}{2}(\omega_0 - \omega)\sigma_z^{\prime} + \tfrac{1}{2}\Omega_{\pi/2}(t) \sigma_x^{\prime} + \tfrac{1}{2}\Omega_{y}(t) \sigma_y^{\prime},
\end{align*}
where the spin lock $H_{y}(t)$ field now acts along the $y^{\prime}$-axis due to the relative phases of $H_{y}(t)$ and the reference frame (a full derivation is provided in Supplementary Note 1). The pulse sequence is illustrated in Fig. \ref{Fig5}(a), where we plot $\Omega_{\pi/2}^{\prime}(t)$ and $\Omega_{y}^{\prime}(t)$ as a function of time. During the $\pi_x/2$ pulses $\Omega_{\pi/2}(t) = \Omega$ for a pulsewidth of $T = \frac{\pi}{2\Omega}$, with $\Omega_{\pi/2}(t) = 0$ otherwise. Conversely, $\Omega_{y}(t) = \Omega$ in between the $\pi_x/2$-pulses, and $\Omega_{y}(t) = 0$ during $\pi_x/2$-pulses. 

The corresponding $S_z^{\prime}$ Bloch vector dynamics are presented in Fig. \ref{Fig5}(b) (see Supplementary movie 11 also). We consider the effect of two drive fields, one that is resonant ($\omega = \omega_0$) and one that is detuned ($\omega \neq \omega_0$) from the Bloch precession. The first $\pi_x/2$-pulse places the Bloch vector in the $x^{\prime}$-$y^{\prime}$ plane, where $S_z^{\prime} = 0$. The spin lock field $\bm{\Omega_{y}^{\prime}}$ then locks the system to the $y^{\prime}$-axis. For the detuned drive field, the locking axis is slightly tilted towards $-\hat{z}$ by $ \bm{\Omega_{\Delta}} = \tfrac{1}{2}(\omega_0 - \omega)\hat{z}$. This is illustrated on the left inset Bloch sphere. The spin lock field $\bm{\Omega_{y}^{\prime}}$ keeps both the resonant and detuned Bloch vectors local to the $y^{\prime}$-axis, so that the final $\pi_x/2$-pulse effectively projects both systems onto -$\hat{z}$. This is illustrated on the right inset Bloch sphere. Note that the $\bm{\Omega_{y}^{\prime}}$ rotation must be faster than the detuning rate $\omega_0 - \omega$ to keep the Bloch vector locked to the $y^{\prime}$-axis. Finally, whilst the spin lock sequence is effective at pinning the Bloch vector to a single axis of the Bloch sphere\cite{10.1093/oso/9780198506348.001.0001, Rizzato2023}, the state cannot vary, which limits the versatility of the technique. 

\subsection{Continuous Concatenated Dynamical Decoupling}\label{CCDD}

Where the spin-lock pins the Bloch precession to the drive frequency, $\omega$, continuous concatenated dynamical decoupling schemes pin the Bloch vector to a Rabi frequency, $\Omega$. This means CCDD schemes operate similarly to a Rabi oscillation, continuously driving the system between $|0\rangle \leftrightarrow |1\rangle$. The locked Rabi frequency enables precise control of the Bloch vector with dramatically enhanced coherence times~\cite{Cai2012, Ramsay2023}, and has been leveraged in high fidelity control schemes~\cite{PhysRevLett.132.223601, Kuno2026}, a suite of quantum sensing applications~\cite{Patrickson2024, Patrickson2025, Stark2017}, and for multipartite entangling operations~\cite{Nnnerich2025}. 

CCDD is typically applied using an amplitude~\cite{Cai2012} or phase~\cite{PhysRevA.96.013850} modulated driving field. Here we focus on phase modulated CCDD, described by the lab frame Hamiltonian, 

\begin{equation}\label{Lab_CCDD}
  H = H_0 + H_{CCDD} = \tfrac{1}{2}\omega_0 \sigma_z + \Omega \cos\left(\omega t - \tfrac{2 \epsilon_m}{\Omega}\sin(\omega_m t)\right)\sigma_x.
\end{equation}

Moving to a reference frame that rotates with the Bloch vector at the drive frequency, $\tfrac{1}{2}\omega \sigma_z$, $H$ transforms into $H^{\prime} = \tfrac{1}{2}(\omega_0 - \omega)\sigma_z^{\prime} + \tfrac{1}{2}\Omega \sigma_x^{\prime} - \epsilon_m \sin(\omega_m t)\sigma_y^{\prime}$ (a full derivation is provided in Supplementary Information note 2). Here $H_{CCDD}$ drives a Rabi oscillation via the DC $x^{\prime}$-component, $\tfrac{1}{2}\Omega \sigma_x^{\prime}$, but also includes an AC $y^{\prime}$-component, $- \epsilon_m \sin(\omega_m t)\sigma_y^{\prime}$. If $\omega_m = \Omega$, this AC modulation locks the Rabi frequency to $\omega_m$, and stabilises the Bloch vector against $(\omega_0 - \omega)\neq0$ drive deviations. This result is not immediately apparent in this first rotating reference frame. However, the interaction can be further simplified by transforming $H^{\prime}$ into a second reference frame that rotates with the Bloch vector Rabi oscillation around $x^{\prime}$, at the AC modulation frequency $\omega_m$. In this case, $H$ reduces to

\begin{equation}\label{Doubly_Rotating_CCDD}
H^{\prime \prime} = \tfrac{1}{2}\epsilon_m \sigma_z^{\prime \prime},
\end{equation}
where we have assumed that $H_{CCDD}$ is resonant with the Bloch precession ($\omega_0 = \omega)$, and the phase modulation is resonant with the Rabi frequency ($\omega_m = \Omega$). In this frame  $H_{CCDD}$ reduces to a DC field along $z^{\prime\prime}$, and is aligned with the state described by, $\bm{S^{\prime\prime}} = (0, 0, 1) = \hat{z}^{\prime \prime}$. 

To understand what the state $\bm{S^{\prime\prime}} = \hat{z}^{\prime \prime}$ represents we must also transform $|\Psi\rangle$ into this frame. Note from Sec. \ref{Rabi_Oscillations}, the first rotating frame states match the lab frame states, so that $|0\rangle^{\prime} = |0\rangle$ and $|1\rangle^{\prime} = |1\rangle$. Transforming into the second rotating frame, we find that,

\begin{align}
    |0^{\prime\prime}\rangle &= U |0^{\prime}\rangle = \cos(\tfrac{1}{2}\omega_mt)|0\rangle + \sin(\tfrac{1}{2}\omega_mt)\sigma_x|0\rangle \nonumber \\ 
    &= \cos(\tfrac{1}{2}\omega_mt)|0\rangle +\sin( \tfrac{1}{2}\omega_mt)|1\rangle,
\end{align}

and,

\begin{align}
    |1^{\prime\prime}\rangle &= U |1^{\prime}\rangle = \cos(\tfrac{1}{2}\omega_mt)|1\rangle + \sin(\tfrac{1}{2}\omega_mt)|1\rangle \nonumber \\ 
    &= \cos(\tfrac{1}{2}\omega_mt)|1\rangle + \sin(\tfrac{1}{2}\omega_mt)|0\rangle.
\end{align}
This means $|0\rangle^{\prime\prime}$ and $|1\rangle^{\prime\prime}$ describe two Rabi oscillations between the lab frame states $|0\rangle$ and $|1\rangle$. $|0\rangle^{\prime\prime}$ and $|1\rangle^{\prime\prime}$ share a common Rabi frequency, $\omega_m$, but are distinguished by a $\pi$ phase shift. Consequently if we apply the CCDD drive $\bm{\Omega_{\mathrm{eff}}^{\prime \prime}} = \tfrac{1}{2}\epsilon_m \sigma_z^{\prime \prime}$ with a Bloch vector prepared in $\bm{S} = \hat{z}^{\prime \prime}$, it will remain in the eigenstate described by the Rabi oscillation $|0\rangle^{\prime\prime}$. If the Bloch vector begins to deviate from this Rabi oscillation, it will instead precess around $\bm{\Omega_{\mathrm{eff}}^{\prime \prime}}$. This response is similar to the spin lock sequence, where detuned systems are effectively pinned to the drive axis.

To illustrate the CCDD dynamics we begin by considering the first rotating reference frame, where the effective field is described by $\bm{\Omega_{CCDD}^{\prime}} = \tfrac{1}{2}(\omega_0 - \omega)\hat{z}^{\prime} + \tfrac{1}{2}\Omega \hat{x}^{\prime} - \epsilon_m \sin(\omega_m t)\hat{y}^{\prime}$. The $x^{\prime}$ and $y^{\prime}$ $\bm{\Omega_{CCDD}^{\prime}}$ components of the sequence are plotted as a function of time in Fig. \ref{Fig5}(c) and (d), respectively. We consider two imperfect drives to demonstrate the robustness of scheme; one detuned from the Bloch precession ($\Delta=\omega - \omega_0\neq0$; $\omega_m = \Omega$, blue in Fig. \ref{Fig5}(c)) and another detuned from the Rabi frequency ( $\Delta=0; \omega_m \neq \Omega$, black in Fig. \ref{Fig5}(c)). 

The $S_z^{\prime}$ trajectories produced by these CCDD drives are presented in Fig. \ref{Fig5}(e), alongside two comparative Rabi oscillations (see Supplementary movie 12 also). A resonant Rabi oscillation ($\omega = \omega_0$, blue) defines the target Rabi frequency and Bloch vector trajectory for the reference frame. A detuned Rabi drive $\omega \neq \omega_0$ (purple) causes the Bloch vector to precess about a tilted axis at an increased generalised Rabi frequency $\Omega_R=\sqrt{\Omega^2 + \Delta^2}$, so that $S_z^{\prime}$ deviates from the resonant trajectory and fails to reach the target state. By comparison, both CCDD drives, detuned from the Bloch precession ($\omega \neq \omega_0$, $\omega_m = \Omega$, orange) and Rabi frequency ($\omega = \omega_0$, $\omega_m \neq \Omega$ yellow), closely track the target Rabi oscillation. A small modulation in $S_z^{\prime}$ is visible when $\omega \neq \omega_0$, $\omega_m = \Omega$ (orange), which arises from a tilted precession axis under the detuned $\omega \neq \omega_0$ field (see Sec. \ref{Spin-Lock}). Bloch sphere trajectories are omitted here, as the overlapping multi-revolution paths obscure rather than clarify the comparison.


These dynamics are also captured in the second reference frame that rotates with the Rabi oscillation defined by $\tfrac{1}{2}\omega_m\hat{x^\prime}$. In Fig. \ref{Fig5}(f) we compare the $S_z^{\prime\prime}$ Bloch components for three conditions in this frame (see Supplementary movie 13 also). First, a CCDD drive resonant with both the Bloch precession and Rabi frequency, $\omega = \omega_0$, $\omega_m = \Omega$, represents the target interaction. This is described by the effective field, $\bm{\Omega_{res}^{\prime \prime}} = \tfrac{1}{2} \epsilon_m \hat{z}^{\prime \prime}$. As the Bloch vector is initialised to $\bm{S_i^{\prime \prime}} = (0, 0, 1) = \hat{z}^{\prime \prime}$, $\bm{\dot{S^{\prime \prime}}} = 0$, and $\bm{S_i^{\prime \prime}}$ represents an eigenstate of $\bm{\Omega_{res}^{\prime \prime}}$ (inset Bloch spheres, blue). We compare the same two imperfect CCDD drives from Figs. \ref{Fig5} (c) - (e). The detuned Rabi frequency condition $\omega = \omega_0$, $\omega_m \neq \Omega$, changes the $\bm{\Omega_{res}^{\prime \prime}}$ axis so that it tilts towards the $x^{\prime \prime}$-axis, $\bm{\Omega_{\mathrm{eff}}^{\prime \prime}} = \tfrac{1}{2}(\Omega - \omega_m)\hat{x}^{\prime \prime} +  \tfrac{1}{2}\epsilon_m \hat{z}^{\prime \prime}$ (inset Bloch spheres, yellow). This causes the Bloch vector to precess around the new $\bm{\Omega_{\mathrm{eff}}^{\prime \prime}}$ axis, producing a small $S_z^{\prime \prime}$ modulation relative to the target trajectory. Note that if $\epsilon_m \gg |(\Omega - \omega_m)|$ then this represents a small perturbation. A detuned Bloch precession $\omega \neq \omega_0$, $\omega_m = \Omega$ introduces two oscillating components into the reference frame (see Supplementary Information note 2 for the derivation), so that $\bm{\Omega_{\mathrm{eff}}^{\prime \prime}} = \tfrac{1}{2}\epsilon_m \hat{z}^{\prime \prime} + \tfrac{1}{2}(\omega-\omega_0)\left(\sin(\omega_mt)\hat{y}^{\prime \prime} + \tfrac{1}{2}\cos(\omega_mt)\hat{z}^{\prime \prime}\right)$. If $\epsilon_m \not\approx \omega_m$, then these perturbations act incoherently on the Bloch vector, simply causing it to oscillate back and forth around $\tfrac{1}{2}\epsilon_m \hat{z}^{\prime \prime}$ (inset Bloch spheres, orange). These conditions have been shown to be forgiving experimentally, making CCDD remarkably robust to inhomogeneous broadening \cite{Ramsay2023, PhysRevLett.132.223601, Kuno2026}.

\section{Coherence Times}\label{Coherence_Times}

A two-level system will lose information about its state over time due to interactions with its environment. This can be understood directly from the Bloch vector, which encodes two types of information: its projection along the $z$-axis, which represents the population of the two energy levels, and its projection in the $x$-$y$ plane, which represents the phase coherence between them. Environmental interactions degrade these two types of information in different ways, leading to three commonly used coherence times.

\subsection{Longitudinal Relaxation Time, $T_1$}\label{T_1}

The longitudinal, or spin-lattice, relaxation time ($T_1$) characterises the timescale over which a two-level system will relax from one of its energy eigenstates, $|0\rangle$, or $|1\rangle$, into thermal equilibrium (which we define below). In any real system, $|0\rangle$ and $|1\rangle$ are not perfectly isolated from their surroundings. They exchange energy with the environment, or bath, via photons, phonons, and/or local fluctuating fields. The strength of this exchange and the bath's temperature determine how quickly population information is lost, and define the $T_1$ relaxation time. On the Bloch sphere, this appears as a motion of the Bloch vector along the $z$-axis towards its equilibrium value, typically either from one energy eigenstate to another, or to a mixed state towards the origin. As such, $T_1$ sets a fundamental upper bound on the coherence time of the system. No quantum operation or algorithm can meaningfully exceed it, since population information will have irreversibly decayed into the thermal background, and with it any encoded quantum state. Note that spin-lock and CCDD lifetime are sometimes described using the symbol $T_{1, \rho}$ \cite{Rizzato2025}. This lifetime is distinct from $T_1$. Its use reflects the long lifetimes achieved by these engineered eigenstates.

To model $T_1$, we must identify the state occupied at thermal equilibrium. How quickly the system relaxes to this thermal state depends on the system-bath coupling, and the difference between the transition frequency $\omega_0$ and the thermal energy $k_BT$. We take an $S=1/2$ electron as an example, where $m_s = -\tfrac{1}{2} = |0\rangle$ and $m_s = \tfrac{1}{2} = |1\rangle$. At room temperature and moderate magnetic field, $\omega_0 \sim \mathrm{GHz}$, and $k_BT \sim \mathrm{THz}$. As $k_BT \gg \omega_0$, the thermal state will approach an equal mixture of $m_s = \pm1/2$ and lie at the origin of the Bloch sphere. At mK temperatures however, $\omega_0 \sim \mathrm{GHz}$, and $k_bT \sim 100's \ \mathrm{MHz}$. As $k_BT < \omega_0$, the system will relax to the lowest energy state available, $m_s = -1/2 = |0\rangle$, corresponding to $\bm{S} = (0, 0, 1) = \hat{z}$ on the Bloch sphere and which we take to be the thermal state in the following models.

\begin{figure*}[h!] 
\centering
\includegraphics[width=1\columnwidth]{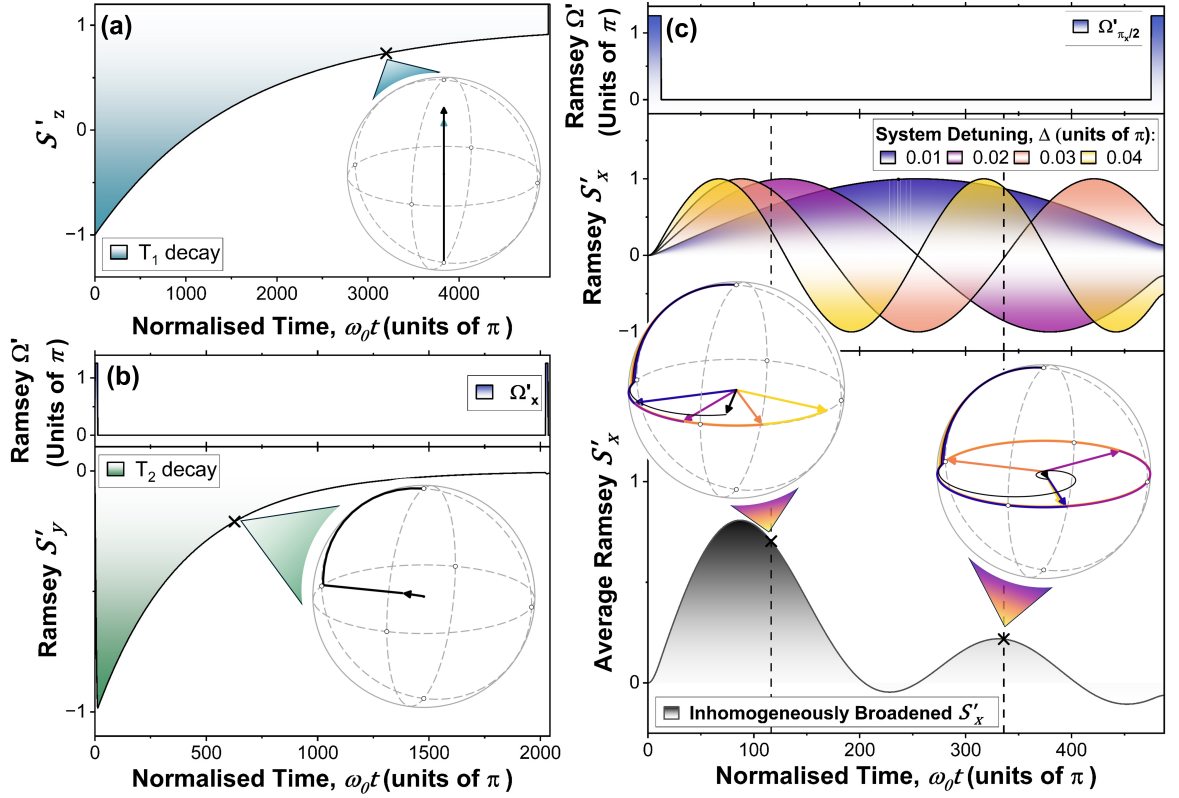}
\caption{Coherence Times. (a) Longitudinal Relaxation, $T_1$. The $S_z$ trajectory of a two-level system undergoing longitudinal relaxation into thermal equilibrium, from $|1\rangle$ ($\bm{S_z = -\hat{z}}$ in the Bloch representation) into $|0\rangle$ ($\bm{S_z = \hat{z}}$). (b) Transverse relaxation, $T_2$. Upper panel, Rotating frame Ramsey pulse sequence. Lower panel, $S_y^{\prime}$ trajectory under transverse relaxation. (c) Inhomogeneous Broadening, $T_2^*$. Upper panel, rotating frame Ramsey pulse sequence. Middle panel, Bloch vector $S_x^{\prime}$ trajectories for a small ensemble of two-level systems. Each system has a different $\omega_0$ Bloch precession frequency, causing it to rotate in the $x^{\prime}$-$y^{\prime}$ plane at a different rate. Consequently the ensemble of Bloch vectors fan out, with diverging $S_x^{\prime}$ components (see inset Bloch sphere). Lower panel, ensemble averaged $S_x^{\prime}$ component. As the ensemble spreads out in the $x^{\prime}$-$y^{\prime}$ plane, the average $S_x^{\prime}$ projection decays (inset Bloch spheres, black trajectory). Although the final $\pi_x/2$-pulse will project each vector onto $S_z$ prior to measurement, the average state of a large ensemble will trend towards 0. The timescale defining this process is known as the inhomogeneous relaxation rate, or $T_2^*$. 
}
\label{Fig6}
\end{figure*}

The mathematical tool used to model decay rates is known as a Lindbladian operator, $L$ \cite{patrickson2026ddtutorial, Nielsen_Chuang_2010}. We provide a brief overview of this formalism in Supplementary note 3, though a full explanation is beyond the scope of this tutorial. For $T_1$ relaxation, it induces an exponential decay of the Bloch vector towards the thermal state. To model the effect we start from an initial state $|\Psi_i\rangle = |1\rangle$, corresponding to $\bm{S_i} = (0, 0, -1) = -\hat{z}$ on the Bloch sphere, and plot the $S_z$ relaxation as a function of time in Fig. \ref{Fig6}(a)  (see also Supplementary movie 14). The inset Bloch sphere plots the full trajectory, where the Bloch vector relaxes from $\bm{S_i} = -\hat{z}$ directly through the origin, towards $\hat{z}$.

\subsection{Transverse Relaxation Time, $T_2$}\label{T_2}
The $T_2$ time describes the loss of information about the phase of a quantum superposition. It originates from complex interactions that couple the system to the local environment, and renders the Bloch vector's trajectory in the $x$-$y$ plane unresolvable. On the Bloch sphere, this appears as a gradual shrinking of the Bloch vector in the $x$-$y$ plane towards the origin.

The effects of $T_2$ dephasing can be observed during the $\tau$ delay period of a Ramsey pulse sequence. To demonstrate this we use the same rotating frame Ramsey Hamiltonian as Fig. \ref{Fig3}, $H^{\prime} = H_{Ramsey}^{\prime} = \tfrac{1}{2}\Omega(t) \sigma_x^{\prime}$, but choose a resonant drive so that $H_{\Delta}^{\prime} = \tfrac{1}{2}(\omega_0 - \omega)\sigma_z = 0$. $T_2$ dephasing is introduced using a Lindbladian operator that causes exponential decay of the Bloch vector in the $x^\prime$-$y^\prime$ plane, describing an irreversible exchange of phase information with the environment. The effect is illustrated in Fig. \ref{Fig6}(b), where we plot the $S_y^{\prime}$ component of the Bloch vector in the lower panel and the corresponding $H_{Ramsey}$ Ramsey pulse sequence in the upper panel (see also Supplementary movie 15). Following the initial $\pi_x/2$-pulse, the $S_y^{\prime}$ Bloch vector component decays exponentially. This trajectory is illustrated on the inset Bloch sphere, where the Bloch vector slowly decays towards the origin where all phase information is lost. Note, this simulation neglects $T_1$ relaxation, and only captures coherence loss from pure dephasing. In practice, $T_2$ dephasing has contributions from both irreversible phase noise and the population relaxation described by $T_1$, and so the two timescales are not independent. In general, $T_2\leq2T_1$, with equality approached only when decoherence is dominated by energy exchange with the environment and there is negligible pure dephasing. Finally we note that $T_2$ relaxation is only applicable to single two-level systems. The dephasing rate for an ensemble is described by inhomogeneous broadening, $T_2^*$.

\subsection{Inhomogeneous Broadening, $T_2^*$}\label{T_2*}
The $T_2^\star$ time describes dephasing that arises from averaging over many realisations of a two-level system. These realisations can be an ensemble of physically distinct systems, or a single system probed repeatedly over time, in which case slow drifts in the local environment play the role of inhomogeneity. In either case, each realisation evolves at a slightly different frequency, causing the average coherence to decay faster than that of any individual system. In the Bloch representation, the vectors describing each two-level system spread out across the $x$-$y$ plane, so that applying an effective field along a single axis will induce different vector rotations across the ensemble. Consequently an average Bloch vector describing the ensemble-averaged state will converge towards the centre of the Bloch sphere, representing an incoherent mixture of states.

The effect of inhomogeneous broadening is illustrated with a Ramsey sequence in Fig. \ref{Fig6}(c). The pulse sequence is presented in the upper panel, and uses the same rotating frame Ramsey Hamiltonian as Fig. \ref{Fig3}, $H^{\prime} = H_{\Delta}^{\prime} + H_{Ramsey}^{\prime} = \tfrac{1}{2}(\omega_0 - \omega)\sigma_z^{\prime} + \tfrac{1}{2}\Omega(t) \sigma_x^{\prime}$. To represent a small ensemble of two-level systems we simulate a range of Bloch precession frequencies, manifesting in different $H_{\Delta}^{\prime} = \tfrac{1}{2}(\omega_0 - \omega)\sigma_z^{\prime}$ terms. The $S_x^{\prime}$ trajectory of each Bloch vector is described in the middle panel of Fig. \ref{Fig6}(c), where each system follows a different evolution  (see also Supplementary movie 16). In the lower panel of Fig. \ref{Fig6}(c) we plot the average $S_x^{\prime}$ component across this ensemble as a function of time, which oscillates whilst decaying towards zero. The oscillation arises from the nonzero average detuning used in the simulation, $\langle \Delta\rangle = 0.025 \pi$, combined with the small number of precession frequencies sampled. Consequently, as individual systems come in and out of coherence with one another, beat frequencies and revivals in the $S_z^{\prime}$ projection become apparent. In the large ensemble limit, this would instead manifest as a smooth decay envelope. Both the ensemble-averaged and individual Bloch vector trajectories are illustrated on the inset Bloch spheres, where the average spirals towards the centre of the sphere.

\subsection{Mitigating Dephasing with Dynamical Decoupling}
Dynamical decoupling protocols extend the observed coherence time of a two-level system by actively suppressing dephasing caused by environmental noise. Rather than reversing the fundamental $T_2$ relaxation process, these sequences apply carefully timed control fields that refocus phase errors arising from slow or quasi-static fluctuations in the system’s precession frequency. On the Bloch sphere, this corresponds to periodically reversing unwanted rotations in the $x-y$ plane, preventing the Bloch vector from spreading due to inhomogeneous or slowly varying detuning, leaving only the intrinsic, irreversible decay processes.

This behaviour can be understood in terms of a noise power spectrum, which describes how environmental fluctuations are distributed in frequency. Dynamical decoupling sequences act as a filter function that selectively suppresses noise within certain frequency bands: the Hahn echo strongly cancels low-frequency, slowly varying noise, while adding additional pulses and shortening $\tau$ with the CPMG sequence will narrow and shift the filter passband to higher frequencies, as phase errors are refocused on shorter timescales. In this picture, the effective coherence time is determined by the overlap between the environmental noise spectrum and the filter function of the applied sequence. High-frequency noise components, which vary on timescales shorter than the pulse spacing, cannot be refocused and therefore continue to contribute to irreversible decoherence. The experimentally measured coherence time thus reflects how well the chosen sequence suppresses the dominant noise in the system. Since the relaxation rate is dependent on the specific control sequence used, the resulting decay time is not equivalent to $T_2$ despite often being labelled as such. Instead, we suggest a sequence-specific notation, writing $T_\mathrm{Rabi}$, $T_\mathrm{CPMG}$, and so forth, to distinguish these quantities clearly.

\section{Conclusions}\label{Conclusion}

To conclude, dynamical decoupling protocols decouple unwanted interactions between a two-level system and its local environment by applying carefully controlled, precise operations that dynamically change the population and phase of the system. Here we express each operation as an effective field vector, and introduce an equation of motion, Eq. \eqref{Eq_of_motion}, to illustrate how a particular field vector will change the phase and population of a given state. This framework describes the two-level system using a Bloch vector, represented on the Bloch sphere. We first demonstrate these concepts by simulating the Bloch vector trajectory of a Rabi oscillation, and illustrate how these dynamics can be simplified in a rotating reference frame. We simulate a Ramsey sequence to introduce the concept of a pulsed control scheme, where the Bloch vector is controlled using precisely timed, discrete operations. These ideas are extended into the pulsed dynamical decoupling Hahn Echo and CPMG protocols, where control pulses periodically reverse the Bloch vector phase to enable self-correction against dephasing environmental interactions. In systems with fast dephasing it can be advantageous to use continuous dynamical decoupling techniques. These protocols apply continuous effective fields to constantly correct against deviations from the target two-level phase and population. We simulate the dynamics of two protocols, the spin lock and continuous concatenated dynamical decoupling. Dynamical decoupling schemes are engineered to extend the useable lifetime of the host quantum system. These timescales are quantified by the longitudinal $T_1$, transverse $T_2$ and inhomogeneous $T_2^*$ lifetimes, which we define in the final section. Together, the concepts introduced in this article provide a complete platform-agnostic framework for interpreting and visualising this range of dynamical decoupling protocols. Each technique is used across the quantum technologies community, with high profile demonstrations spanning quantum sensing \cite{Rovny2025}, quantum memory \cite{PhysRevX.9.031045} and entanglement operations \cite{Nnnerich2025}. Despite their established status, they continue to push the state-of-the-art, representing a critical tool in the quantum engineer's toolbox. 

\section{Modelling}

The two-level spin dynamics are numerically obtained by integrating the von-Neumann equation, $\dot\rho = -i[H(t),\rho]$, with optional dissipative contributions included via Lindblad terms of the form $L\rho L^\dagger - \frac{1}{2}\{L^\dagger L,\rho\}$. The Hamiltonian is constructed using combinations of Pauli operators with time-dependent coefficients. At each time step, the variation of the density matrix over time is evaluated and implemented into a high accuracy ODE solver. The resulting state is then projected onto the Pauli basis to extract expectation values. While the Hamiltonian itself is decomposed into its composite Pauli operators. These results are then saved, to be used for displaying the results on the Bloch sphere.

\section{Acknowledgements}
We would like to express our sincere gratitude to Andrew J. Ramsay for sharing his expertise in spin dynamics, and for the many fruitful discussions that have arisen from his input. We acknowledge and are grateful to the EPSRC quantum career development grant EP/W028301/1, the EPSRC Standard Research grant EP/Z534250/1, and an EPSRC iCASE in partnership with Oxford Instruments Plasma Technology (I.J.L. and C.J.P.). Ion implantation was performed by Keith Heasman and Julian Fletcher at the University of Surrey Ion Beam Centre.

\providecommand{\noopsort}[1]{}\providecommand{\singleletter}[1]{#1}%


\providecommand{\noopsort}[1]{}\providecommand{\singleletter}[1]{#1}%
\begin{thebibliography}{10}

\bibitem{Murado2025}
M. Murado\u{g}lu et al., 2025 Quantum-assured magnetic navigation achieves positioning accuracy better than a strategic-grade ins in airborne and ground-based field trials, DOI: https://doi.org/10.48550/arXiv.2504.08167


\bibitem{Zhong2020}
H. S. Zhong et al., 2020 Quantum computational advantage using photons. Science 370, 1460 - 1463, DOI: https://doi.org/10.1126/science.abe8770

\bibitem{Yin2017}
J. Yin et al., 2017 Satellite-based entanglement distribution over 1200 kilometers. Science 356, 1140 - 1144, DOI: https://doi.org/10.1126/science.aan3211

\bibitem{Pittaluga2025}
M. Pittaluga et al., 2025 Long-distance coherent quantum communications in deployed telecom networks, Nature 640, 911 - 917, DOI:https://doi.org/10.1038/s41586-025-08801-w

\bibitem{Dyte2025}
H. E. Dyte, S. Manna, S. F. Covre da Silva, A. Rastelli, and E. A. Chekhovich. 2026 Storing quantum coherence in a quantum dot nuclear spin ensemble for over 100 milliseconds, Nature Communications 17, 239, DOI: 
https://doi.org/10.1038/s41467-025-66948-6

\bibitem{Zaporski2023}
L. Zaporski et al., 2023 Ideal refocusing of an optically active spin qubit under strong hyperfine interactions, Nature Nanotechnology 18, 257- 263, DOI: https://doi.org/10.1038/s41565-022-01282-2

\bibitem{PhysRevX.9.031045}
C. E. Bradley, J. Randall, M. H. Abobeih, R.C. Berrevoets, M.J. Degen, M.A. Bakker, M. Markham, D. J. Twitchen, and T. H. Taminiau, 2019 A ten-qubit solid-state spin register with quantum memory up to one minute, Physical Review X 9, 031045, DOI: https://doi.org/10.1103/PhysRevX.9.031045

\bibitem{Rizzato2025}
R. Rizzato, A. A. Hidalgo, L. Nie, E. Blundo, N. R. von Grafenstein, J. J. Finley, and D. B. Bucher, 2025 Quantum sensing with spin defects in boron nitride nanotubes, Nature Communications 16, 11333, https://doi.org/10.1038/s41467-025-67538-2

\bibitem{PhysRevLett.132.223601}
A. Salhov, Q. Cao, J. Cai, A. Retzker, F. Jelezko, and G. Genov, 2024 Protecting quantum information via destructive interference of correlated noise, Physical Review Letters 132, 223601, DOI: https://doi.org/10.1103/PhysRevLett.132.223601

\bibitem{PhysRevLett.121.220502}
B. Pokharel, N. Anand, B. Fortman, and D.A. Lidar, 2018 Demonstration of fidelity improvement using dynamical decoupling with superconducting qubits, Physical Review Letters 121, 220502, DOI: https://doi.org/10.1103/PhysRevLett.121.220502

\bibitem{Drmota2023}
P. Drmota et al., 2023 Robust quantum memory in a trapped-ion quantum network node, Physical Review Letters 130, 090803, DOI: https://doi.org/10.1103/PhysRevLett.130.090803

\bibitem{PhysRevApplied.11.014017}
D.J. Egger, M. Ganzhorn, G. Salis, A. Fuhrer, P. M{\"u}ller, P.Kl. Barkoutsos, N.~Moll, I.~Tavernelli, and S.~Filipp, 2019 Entanglement generation in superconducting qubits using holonomic operations, Physical Review Applied 11, 014017, DOI: https://doi.org/10.1103/PhysRevApplied.11.014017

\bibitem{Nnnerich2025}
M. Nünnerich, D. Cohen, P. Barthel, P. H. Huber, D. Niroomand, A. Retzker, and C. Wunderlic, 2025 Fast, robust, and laser-free universal entangling gates for trapped-ion quantum computing, Physical Review X 15, 021079, DOI: https://doi.org/10.1103/PhysRevX.15.021079

\bibitem{PhysRevLett.111.067601}
P. London et al., 2013 Detecting and polarizing nuclear spins with double resonance on a single electron spin, Physical Review Letters 111, 067601, DOI: https://doi.org/10.1103/PhysRevLett.111.067601

\bibitem{Rovny2025}
J. Rovny, S. Kolkowitz, and N. P. de Leon, 2025 Multi-qubit nanoscale sensing with entanglement as a resource, Nature 647, 876–882, https://doi.org/10.1038/s41586-025-09760-y

\bibitem{vonLpke2024}
U. von Lüpke, I.C. Rodrigues, Y. Yang, M. Fadel, and Y. Chu, 2024 Engineering multimode interactions in circuit quantum acoustodynamics, Nature Physics 20, 564–570, https://doi.org/10.1038/s41567-023-02377-w

\bibitem{Shulman2012}
M. D. Shulman, O. E. Dial, S. P. Harvey, H. Bluhm, V. Umansky, and A. Yacoby, 2012 Demonstration of entanglement of electrostatically coupled singlet-triplet qubits, Science 336, 202-205, DOI: https://doi.org/10.1126/science.1217692

\bibitem{Degen2017}
C. L. Degen, F. Reinhard, and P. Cappellaro, 2017 Quantum sensing, Reviews of Modern Physics 89, 035002, DOI: https://doi.org/10.1103/RevModPhys.89.035002

\bibitem{patrickson2026ddtutorial}
C. Patrickson, I. Bailey, A. Orazzo, L. Dellantonio., I. J. Luxmoore, 2026 Tutorial: Interpreting spin dynamics in dynamical decoupling schemes, GitHub repository, https://github.com/Charlie-Patrickson/Tutorial-Interpreting-Spin-Dynamics-in-Dynamical-Decoupling-Schemes/

\bibitem{patrickson2026spinDynamicsTutorial}
C. Patrickson, I. Bailey, A. Orazzo, L. Dellantonio., I. J. Luxmoore, 2026 Tutorial: Interpreting spin dynamics in dynamical decoupling schemes, Online preprint/tutorial, \\ 
https://charlie-patrickson.github.io/Tutorial-Interpreting-Spin-Dynamics-in-Dynamical-Decoupling-Schemes/

\bibitem{Nielsen_Chuang_2010}
M. A. Nielsen and I. L. Chuang. 2010 Quantum Computation and Quantum Information: 10th Anniversary Edition, Cambridge University Press, DOI: https://doi.org/10.1017/CBO9780511976667

\bibitem{Tetienne2012}
J. P. Tetienne, L. Rondin, P. Spinicelli, M. Chipaux, T. Debuisschert, J. F. Roch, and V. Jacques, 2012 Magnetic-field-dependent photodynamics of single nv defects in diamond: an application to qualitative all-optical magnetic imaging, New Journal of Physics 14, 103033, DOI: https://doi.org/10.1088/1367-2630/14/10/103033

\bibitem{10.1093/oso/9780198506348.001.0001}
A. Schweiger and G. Jeschke, 2001 Principles of Pulse Electron Paramagnetic Resonance, Oxford University Press, DOI: https://doi.org/10.1002/jctb.936

\bibitem{Debroux2021}
R. Debrouxet al., Quantum control of the tin-vacancy spin qubit in diamond, Physical Review X 11, 041041, DOI: https://doi.org/10.1103/PhysRevX.11.041041

\bibitem{Stern2024}
Hannah L. Stern et al., 2024 A quantum coherent spin in hexagonal boron nitride at ambient conditions, Nature Materials 23, 1379–1385, DOI: https://doi.org/10.1038/s41563-024-01887-z

\bibitem{Hahn1950}
E. L. Hahn, 1950 Spin echoes, Physical Review 80, 580, DOI: https://doi.org/10.1103/PhysRev.80.580

\bibitem{Shofer2025}
N. Shofer et al., 2025 Tuning the coherent interaction of an electron qubit and a nuclear magnon, Physical Review X 15, 021004, DOI: https://doi.org/10.1103/PhysRevX.15.021004

\bibitem{Choi2020}
J. Choi, H. Zhou, H.S. Knowles, R. Landig, S. Choi, and M.D. Lukin. 2020 Robust dynamic hamiltonian engineering of many-body spin systems, Physical Review X 10, 031002, DOI: https://doi.org/10.1103/PhysRevX.10.031002

\bibitem{Biercuk2009}
M.J. Biercuk, H. Uys, A.P. VanDevender, N. Shiga, W.M. Itano, and J.J. Bollinger, 2009 Optimized dynamical decoupling in a model quantum memory, Nature 458, 996–1000, DOI: https://doi.org/10.1038/nature07951

\bibitem{Bodey2019}
J. H. Bodey et al., 2019 Optical spin locking of a solid-state qubit, npj Quantum Information 5, 95, DOI: https://doi.org/10.1038/s41534-019-0206-3 

\bibitem{Rizzato2023}
R. Rizzato et al., 2023 Extending the coherence of spin defects in hbn enables advanced qubit control and quantum sensing, Nature Communication 14, 5089, DOI: https://doi.org/10.1038/s41467-023-40473-w

\bibitem{Cai2012}
J.~M. Cai, B.~Naydenov, R.~Pfeiffer, L.~P. McGuinness, K.~D. Jahnke, F.~Jelezko, M.~B. Plenio, and A.~Retzker, 2012 Robust dynamical decoupling with concatenated continuous driving, New Journal of Physics 14, 113023, DOI: https://doi.org/10.1088/1367-2630/14/11/113023

\bibitem{Ramsay2023}
A.J. Ramsay, R. Hekmati, C.J. Patrickson, S. Baber, D.R.M. Arvidsson-Shukur, A.J. Bennett, and I.J. Luxmoore, 2023 Coherence protection of spin qubits in hexagonal boron nitride, Nature Communications 14, 461, DOI: https://doi.org/10.1038/s41467-023-36196-7

\bibitem{Kuno2026}
T. Kuno et al., 2026 Concatenated continuous driving of silicon qubit by amplitude and phase modulation, Physical Review B 113, 195303, DOI: https://doi.org/10.1103/f2kd-x628

\bibitem{Patrickson2024}
C.J. Patrickson, S. Baber, B.B. Gaál, A.J. Ramsay, and I.J. Luxmoore, 2024 High frequency magnetometry with an ensemble of spin qubits in hexagonal boron nitride, npj Quantum Information 10, 5, DOI
https://doi.org/10.1038/s41534-023-00796-4

\bibitem{Patrickson2025}
C.J. Patrickson, V. Haemmerli, S. Guo, A.J. Ramsay, and I.J. Luxmoore, 2025 Microwave quantum heterodyne sensing using a continuous concatenated dynamical decoupling protocol, Nature Communications 16, 4380, DOI
https://doi.org/10.1038/s41467-025-59148-9

\bibitem{Stark2017}
A. Stark, N. Aharon, T. Unden, D. Louzon, A. Huck, A. Retzker, U.L. Andersen, and F. Jelezko, 2017 Narrow-bandwidth sensing of high-frequency fields with continuous dynamical decoupling, Nature Communications 8, 1105, DOI
https://doi.org/10.1038/s41467-017-01159-2

\bibitem{PhysRevA.96.013850}
D.~Farfurnik, N.~Aharon, I.~Cohen, Y.~Hovav, A.~Retzker, and N.~Bar-Gill, 2017 Experimental realization of time-dependent phase-modulated continuous dynamical decoupling, Physical Review A 96, 013850, DOI: https://doi.org/10.1103/PhysRevA.96.013850

\end{thebibliography}
\end{document}


\articletype{Tutorial} 

\title{Supplementary Information: A Quantum Dynamics Tutorial: Visualising Dynamical Decoupling Sequences on the Bloch Sphere}


\author{Charlie J. Patrickson$^{1, *}$\orcid{0000-0003-0550-6396}, Isaac Bailey$^2$, Antonio Orazzo$^2$, Luca Dellantonio, $^2$\orcid{0000-0000-0000-0000} and Isaac J. Luxmoore$^{1}$\orcid{0000-0002-2650-0842}}

\affil{$^1$Department of Engineering, University of Exeter, EX4 4QF, UK} 
\affil{$^2$Department of Physics and Astronomy, University of Exeter, EX4 4QF, UK} 

\affil{$^*$Author to whom any correspondence should be addressed.}

\email{cp728@exeter.ac.uk}

\keywords{Dynamical Decoupling, Tutorial, Spin Dynamics}

\section{Supplementary Note 1: Rabi, Ramsey, Hahn Echo, CPMG and Spin Lock Rotating Reference Frame Transformations}\label{Reference_Frame_Transformations}

Throughout the main text we present Bloch vector trajectories in a rotating reference frame that tracks with the Bloch precession term. These trajectories are calculated in the interaction picture, whereby the Hamiltonian governing the interaction is transformed into the target reference frame. Here we provide a detailed derivation of these transformations for the Rabi, Ramsey, Hahn echo, CPMG and spin lock sequences. Each can be derived from the lab frame Hamiltonian, $H$, 
\begin{eqnarray}
H = H_0 + H_{Drive} = \tfrac{1}{2} \omega_0\sigma_z + \Omega (t) \cos(\omega t + \phi)\sigma_x.
\end{eqnarray}
$H_0 =  \tfrac{1}{2}\omega_0 \sigma_z$ describes the Bloch precession, and $H_{Drive} = \Omega (t)\cos(\omega t+ \phi)\sigma_x$ describes the continuous or pulsed drive used in the corresponding control sequence. To transform $H$ into a reference frame that rotates with the Bloch precession, $\tfrac{1}{2} \omega_0\sigma_z $, we use the matrix $U = e^{i\tfrac{1}{2}\omega t \sigma_z}$, which obeys the identity $e^{i\tfrac{1}{2}\omega t \sigma_z} = \cos(\tfrac{1}{2}\omega t) + i \sin(\tfrac{1}{2}\omega t)\sigma_z$. The transformed Hamiltonian, $H^{\prime}$, is calculated using 
\begin{eqnarray}
H^{\prime} = UHU^{\dagger} + i\dot{U}U^{\dagger}.
\end{eqnarray}
Substituting in $H$ and $U$, 
\begin{eqnarray}
H^{\prime} = e^{i\tfrac{1}{2}\omega t \sigma_z} \left( \tfrac{1}{2}\omega_0 \sigma_z + \Omega (t) \cos(\omega t+ \phi) \sigma_x \right) e^{-i\tfrac{1}{2}\omega t \sigma_z} + i \tfrac{d}{dt}e^{i\tfrac{1}{2}\omega t \sigma_z} e^{-i\tfrac{1}{2}\omega t \sigma_z}.
\end{eqnarray}
The time derivative $i \tfrac{d}{dt}e^{i\tfrac{1}{2}\omega t \sigma_z} e^{-i\tfrac{1}{2}\omega t \sigma_z}$ conveniently reduces to $- \tfrac{1}{2}\omega_0 \sigma_z$. Substituting in $U = \cos(\tfrac{1}{2}\omega t) + i \sin(\tfrac{1}{2}\omega t)\sigma_z$,
\begin{eqnarray}
H^{\prime} =  \left(\cos(\tfrac{1}{2}\omega t) + i \sin(\tfrac{1}{2}\omega t)\sigma_z\right) \left( \tfrac{1}{2}\omega_0 \sigma_z + \Omega (t) \cos(\omega t+ \phi) \sigma_x \right)\left(\cos(\tfrac{1}{2}\omega t) - i \sin(\tfrac{1}{2}\omega t)\sigma_z\right) - \tfrac{1}{2}\omega_0 \sigma_z.
\end{eqnarray}
Multiplying through and retaining the order of the Pauli matrices, we arrive at,
\begin{eqnarray}
H^{\prime} = \left(\cos^2(\tfrac{1}{2}\omega t) + \sin^2(\tfrac{1}{2}\omega t)\sigma_z^2\right) \tfrac{1}{2}\omega_0\sigma_z + \left(\cos^2(\tfrac{1}{2}\omega t)\sigma_x + \sin^2(\tfrac{1}{2}\omega t)\sigma_z\sigma_x\sigma_z\right)\Omega (t) \cos(\omega t+ \phi) \cdots\nonumber \\ \cdots + \left( i\sin(\tfrac{1}{2}\omega t)\cos(\tfrac{1}{2}\omega t)\sigma_z\sigma_x - i\sin(\tfrac{1}{2}\omega t)\cos(\tfrac{1}{2}\omega t)\sigma_x\sigma_z \right)\Omega (t) \cos(\omega t+ \phi) -\tfrac{1}{2}\omega_0 \sigma_z.
\end{eqnarray}
Applying the identities $\cos^2(\theta) + \sin^2(\theta) = 1$, $\sigma_z\sigma_x = i\sigma_y$, $\sigma_x\sigma_z = -i\sigma_y$ and $\sigma_z\sigma_x\sigma_z = -i\sigma_x$, we find
\begin{align}
H^{\prime} = \tfrac{1}{2}\omega \sigma_z - \tfrac{1}{2}\omega_0\sigma_z + \left(\cos^2(\tfrac{1}{2}\omega_0t)- \sin^2(\tfrac{1}{2}\omega_0t)\right)\Omega (t) \cos(\omega_0t + \phi)\sigma_x \nonumber \cdots \\ \cdots + \left(-2\sin(\tfrac{1}{2}\omega_0t)\cos(\tfrac{1}{2}\omega_0t)\sigma_y \right)\Omega (t) \cos(\omega t + \phi)),
\end{align}
which can be further simplified with the trigonometric identities, $\sin(2\theta) = 2\sin(\theta)\cos(\theta)$ and $\cos(2\theta) = \cos^2(\theta) - \sin^2(\theta)$;
\begin{eqnarray}
H^{\prime} = \tfrac{1}{2}(\omega- \omega_0)\sigma_z+\cos(\omega_0t)\Omega (t) \cos(\omega t + \phi)\sigma_x - \sin(\omega_0t)\Omega (t) \cos(\omega t+ \phi)\sigma_y.
\end{eqnarray}
To remove the time dependence of the trigonometric terms, we first use $\cos(\theta)\cos(\phi) = \tfrac{1}{2}(\cos(\theta - \phi) + \cos(\theta + \phi))$ and $\sin(\theta)cos(\phi) = \tfrac{1}{2} (\sin(\theta + \phi) + \sin(\theta - \phi))$;
\begin{align}\label{Frame_Transformation_pickup}
H^{\prime} = \tfrac{1}{2}(\omega- \omega_0)\sigma_z+\tfrac{1}{2}\Omega (t)\left(\cos(\omega_0t - \omega t - \phi) + \cos(\omega_0t + \omega t + \phi)\right)\sigma_x \cdots \nonumber \\ 
\cdots - \tfrac{1}{2}\Omega (t)\left(\sin(\omega_0t + \omega t + \phi) + \sin(\omega_0t - \omega t - \phi)\right)\sigma_y.
\end{align}

At this point our choice of $\phi$ will impact how we apply this Hamiltonian. For control protocols that apply a field along the the $x$ ($y$) direction, $\phi = 0$ ($\phi = \tfrac{\pi}{2}$). We assume that, whilst some detuning may be present, $\omega$ and $\omega_0$ are comparable in magnitude, so that $\tfrac{1}{2}\Omega (t)(\cos(\omega_0t - \omega t - \phi))\approx\tfrac{1}{2}\Omega (t)\cos(\phi)$ and $\tfrac{1}{2}\Omega (t)(\sin(\omega_0t - \omega t - \phi))\approx\tfrac{1}{2}\Omega (t)\sin(\phi)$. Applying these approximations we find,
\begin{eqnarray}
H^{\prime}_{\phi=0}
&\approx\tfrac{1}{2}(\omega- \omega_0)\sigma_z+\tfrac{1}{2}\Omega (t)\left(1 + \cos(\omega_0 + \omega)t\right)\sigma_x - \tfrac{1}{2}\Omega (t)\left(\sin(\omega_0 + \omega)t\right)\sigma_y \\
H^{\prime}_{\phi=\tfrac{\pi}{2}}
&\approx\tfrac{1}{2}(\omega- \omega_0)\sigma_z+\tfrac{1}{2}\Omega (t)\cos((\omega_0 + \omega)t + \tfrac{\pi}{2})\sigma_x - \tfrac{1}{2}\Omega (t)\left(1+ \sin((\omega_0 + \omega)t + \tfrac{\pi}{2})\right)\sigma_y
\end{eqnarray}
Applying the rotating wave approximation we disregard the fast oscillating $(\omega_0 + \omega)$ terms, to arrive at,
\begin{eqnarray}
H^{\prime}_{\phi=0} = \tfrac{1}{2}(\omega- \omega_0)\sigma_z+\tfrac{1}{2}\Omega (t) \sigma_x \\
H^{\prime}_{\phi=\tfrac{\pi}{2}} = \tfrac{1}{2}(\omega- \omega_0)\sigma_z-\tfrac{1}{2}\Omega (t) \sigma_y
\end{eqnarray}
Here we see that including a drive phase $\phi$ allows us to apply pulses or drive fields along either the $x$ or $y$ axes. This is a direct consequence of the Bloch vector's constant precession; although the drive is applied along the $x$ axis in the lab, the vector's intrinsic rotation in the $xy$-plane allows us to align precisely timed pulses with the Bloch vector's $y$-axis. The time dependency of $\Omega(t)$ defines the timing of these pulses, with control protocols periodically switching between $\Omega = 0$ and $\Omega \neq 0$ to drive coherent rotations of the Bloch vector, as described in the main text. Finally we note that for the Rabi oscillation in Fig. 2 we selected a resonant drive, so that $\omega = \omega_0$, and $H^{\prime} = \tfrac{1}{2}\Omega \sigma_x$. 

\section{Supplementary Note 2: Reference Frame Transformations of a Continuous Concatenated Dynamical Decoupling Drive}\label{CCDD_Transformations}

In Sec. 7.2 of the main text we introduce the continuous concatenated dynamical decoupling drive. The CCDD Hamiltonian is often transformed into a doubly rotating reference frame to reduce the time dependency. In this section we provide a step-by-step derivation of this transformation. The lab frame CCDD Hamiltonian, which includes the Bloch precession term, $H_0$, and the CCDD drive field, $H_{CCDD}$, takes the form,
\begin{equation}\label{Lab_CCDD}
  H = H_0 + H_{CCDD} = \tfrac{1}{2}\omega_0 \sigma_z + \Omega \cos\left(\omega t - \tfrac{2 \epsilon_m}{\Omega}\sin(\omega_m t)\right).
\end{equation}
We begin by noting the similarity between $H_{CCDD}$ and the drive Hamiltonian used in the previous frame transformations (see Supplementary Note \ref{Reference_Frame_Transformations}), where $H_{Drive} = \Omega (t) \cos(\omega t + \phi)\sigma_x$; if we substitute $\phi = - \tfrac{2 \epsilon_m}{\Omega}\sin(\omega_m t)$ into $H_{Drive}$, we recover $H_{CCDD}$. To avoid repetition, we therefore continue the CCDD transformation from Eq. \ref{Frame_Transformation_pickup}, and substitute $\phi = - \tfrac{2 \epsilon_m}{\Omega}\sin(\omega_m t)$, 
\begin{eqnarray}\label
HH^{\prime} = \tfrac{1}{2}(\omega- \omega_0)\sigma_z+\tfrac{1}{2}\Omega (t)\left(\cos(\omega_0t - \omega t + \tfrac{2 \epsilon_m}{\Omega}\sin(\omega_m t)) + \cos(\omega_0t + \omega t -\tfrac{2 \epsilon_m}{\Omega}\sin(\omega_m t))\right)\sigma_x \cdots \nonumber \\ \cdots - \tfrac{1}{2}\Omega (t)\left(\sin(\omega_0t + \omega t - \tfrac{2 \epsilon_m}{\Omega}\sin(\omega_m t)) + \sin(\omega_0t - \omega t + \tfrac{2 \epsilon_m}{\Omega}\sin(\omega_m t))\right)\sigma_y.
\end{eqnarray}
As before, we assume that $\omega$ and $\omega_0$ are comparable in magnitude, so that $\tfrac{1}{2}\Omega (t)(\cos(\omega_0t - \omega t + \tfrac{2 \epsilon_m}{\Omega}\sin(\omega_m t))))\approx\tfrac{1}{2}\Omega (t)\cos(\tfrac{2 \epsilon_m}{\Omega}\sin(\omega_m t)))$ and $\tfrac{1}{2}\Omega (t)(\sin(\omega_0t - \omega t + \tfrac{2 \epsilon_m}{\Omega}\sin(\omega_m t))))\approx\tfrac{1}{2}\Omega (t)\sin(\tfrac{2 \epsilon_m}{\Omega}\sin(\omega_m t)))$. In most experiments $\epsilon_m \ll \Omega$ so that we can use the small angle approximations, $\sin(\phi) \approx \phi$, and $\cos(\phi) \approx 1$. Applying these approximations, we find, 
\begin{eqnarray}
H^{\prime} = \tfrac{1}{2}(\omega- \omega_0)\sigma_z+\tfrac{1}{2}\Omega (t)\left(1 + \cos((\omega_0 + \omega)t -\tfrac{2 \epsilon_m}{\Omega}\sin(\omega_m t))\right)\sigma_x \cdots \nonumber \\ \cdots - \tfrac{1}{2}\Omega (t)\left(\sin((\omega_0 + \omega)t - \tfrac{2 \epsilon_m}{\Omega}\sin(\omega_m t)) + \tfrac{2 \epsilon_m}{\Omega}\sin(\omega_m t))\right)\sigma_y.
\end{eqnarray}
As CCDD experiments apply continuous fields rather than pulses we can assume the amplitude $\Omega$ to be constant in time. Applying the rotating wave approximation we disregard the fast oscillating $\omega_0 + \omega$ terms, so that
\begin{eqnarray}
H^{\prime} = (\omega_0 - \omega)\sigma_z^{\prime} + \tfrac{1}{2}\Omega \sigma_x^{\prime} - \epsilon_m \sin(\omega_m t)\sigma_y^{\prime}.
\end{eqnarray}
Moving to a reference frame that rotates around the $x^{\prime}$-axis at the CCDD modulation frequency, $\omega_m$ , so that now $U = e^{i\tfrac{1}{2}\omega_m t \sigma_x^{\prime}}$,
\begin{eqnarray}
H^{\prime} = U\left((\omega_0 - \omega)\sigma_z^{\prime} + \tfrac{1}{2}\Omega \sigma_x^{\prime} - \epsilon_m \sin(\omega_m t)\sigma_y^{\prime}\right)U^{\dagger} + i\dot{U}U^{\dagger}.
\end{eqnarray}
Applying the same identity,  $U = e^{i\tfrac{1}{2}\omega_m t \sigma_x} = \cos(\tfrac{1}{2}\omega_m t) + i \sin(\tfrac{1}{2}\omega_m t)\sigma_x$, 
\begin{eqnarray}
H^{\prime\prime} = (\cos(\tfrac{1}{2}\omega_m t) + i \sin(\tfrac{1}{2}\omega_m t)\sigma_x^{\prime\prime})\big(\tfrac{1}{2}(\omega_0 - \omega)\sigma_z^{\prime\prime} \cdots \nonumber \\ \cdots + \tfrac{1}{2}\Omega \sigma_x^{\prime\prime} - \epsilon_m \sin(\omega_m t)\sigma_y^{\prime\prime}\big)(\cos(\tfrac{1}{2}\omega_m t) - i \sin(\tfrac{1}{2}\omega_m t)\sigma_x^{\prime\prime}) - \tfrac{1}{2}\omega_m \sigma_x^{\prime\prime}.
\end{eqnarray}
where the double prime denotes the second rotating reference frame. For simplicity we separate this into two components. First we consider the CCDD drive field, 
\begin{eqnarray}
H_{CCDD}^{\prime\prime} = (\cos(\tfrac{1}{2}\omega_m t) + i \sin(\tfrac{1}{2}\omega_m t))\sigma_x^{\prime\prime}\left( - \epsilon_m \sin(\omega_m t)\sigma_y^{\prime\prime}\right)(\cos(\tfrac{1}{2}\omega_m t) - i \sin(\tfrac{1}{2}\omega_m t))\sigma_x^{\prime\prime}.
\end{eqnarray}
Multiplying out we arrive at,
\begin{eqnarray}
H_{CCDD}^{\prime\prime} = - \epsilon_m \sin(\omega_m t)(\cos^2(\tfrac{1}{2}\omega_m t)\sigma_y^{\prime\prime} - i^2 \sin^2(\tfrac{1}{2}\omega_m t)\sigma_x^{\prime\prime}\sigma_y^{\prime\prime}\sigma_x^{\prime\prime} \cdots \nonumber \\ \cdots - i\cos(\tfrac{1}{2}\omega_m t)\sin(\tfrac{1}{2}\omega_m t))\sigma_y^{\prime\prime}\sigma_x^{\prime\prime} + i\cos(\tfrac{1}{2}\omega_m t)\sin(\tfrac{1}{2}\omega_m t))\sigma_x^{\prime\prime}\sigma_y^{\prime\prime}.
\end{eqnarray}
Here we apply the Pauli matrix transformations $\sigma_x\sigma_y = i\sigma_z$, $\sigma_y\sigma_x = -i\sigma_z$ and $\sigma_x\sigma_y\sigma_x = -\sigma_y$, and the trigonometric identities $\sin(2\theta) = 2\sin(\theta)\cos(\theta)$ and $\cos(2\theta) = \cos^2(\theta) - \sin^2(\theta)$, to arrive at,
\begin{eqnarray}
H_{CCDD}^{\prime\prime} = - \epsilon_m \sin(\omega_m t)(\cos(\omega_m t)\sigma_y^{\prime\prime}- \sin(\omega_m t))\sigma_z^{\prime\prime}.
\end{eqnarray}
Now applying the trigonometric identities $\sin(\theta)\cos(\phi) = \tfrac{1}{2}(\sin(\theta + \phi) + \sin(\theta - \phi))$ and $\sin(\theta)\sin(\phi) = \tfrac{1}{2}(\cos(\theta - \phi) - \cos(\theta + \phi))$,
\begin{align}
H_{CCDD}^{\prime\prime} &= - \tfrac{1}{2}\epsilon_m (\sin(2\omega_m t) + \sin(0))\sigma_y^{\prime\prime}- (\cos(0) + \cos(2\omega_m t))\sigma_z^{\prime\prime} \\
&= \tfrac{1}{2}\epsilon_m\sigma_z^{\prime\prime}.
\end{align}
where in the final step we have applied the RWA. Returning to the remaining Bloch precession detuning term, $\tfrac{1}{2}(\omega_0 - \omega)\sigma_z^{\prime\prime}$ and the Rabi detuning term, $\tfrac{1}{2}(\Omega -\omega_m)\sigma_x^{\prime\prime}$;
\begin{eqnarray}
H_{CCDD \ Detuning}^{\prime\prime} = \tfrac{1}{2}(\Omega -\omega_m)\sigma_x^{\prime\prime} + \tfrac{1}{2}(\omega_0 - \omega)\left(\cos^2(\tfrac{1}{2}\omega_m t)\sigma_z^{\prime\prime} + \sin^2(\tfrac{1}{2}\omega_m t) \sigma_x^{\prime\prime}\sigma_z^{\prime\prime}\sigma_x^{\prime\prime} \right) + \cdots \nonumber \\ \cdots \tfrac{1}{2}(\omega_0 - \omega)\left(i \sin(\tfrac{1}{2}\omega_m t)\cos(\tfrac{1}{2}\omega_m t)\sigma_x^{\prime \prime} \sigma_z^{\prime \prime} -i \sin(\tfrac{1}{2}\omega_m t)\cos(\tfrac{1}{2}\omega_m t)\sigma_z^{\prime \prime} \sigma_x^{\prime \prime} \right)
\end{eqnarray}
Here we apply the Pauli matrix transformations $\sigma_z\sigma_x = i\sigma_y$, $\sigma_x\sigma_z = -i\sigma_y$ and $\sigma_x\sigma_z\sigma_x = i\sigma_z$, and the trigonometric identities $\sin(2\theta) = 2\sin(\theta)\cos(\theta)$ and $\cos(2\theta) = \cos^2(\theta) - \sin^2(\theta)$, to arrive at,
\begin{eqnarray}
H_{CCDD \ Detuning}^{\prime\prime} = \tfrac{1}{2}(\Omega -\omega_m)\sigma_x^{\prime\prime} + \tfrac{1}{2}(\omega_0 - \omega)\cos(\omega_m t)\sigma_z^{\prime\prime} + \tfrac{1}{2}(\omega_0 - \omega)(\sin(\omega_m t) \sigma_y^{\prime \prime}).
\end{eqnarray}

\section{Supplementary Note 3: Simulations}\label{Simulations}

The two-level spin dynamics are numerically obtained by integrating the von-Neumann equation, $\dot\rho = - i[H(t),\rho]$, with optional dissipative contributions included via Lindblad terms of the form $L\rho L^\dagger - \frac{1}{2}\{L^\dagger L,\rho\}$. The Hamiltonian is constructed using combinations of Pauli operators with time-dependent coefficients. At each time step, the variation of the density matrix over time is evaluated and implemented into a high accuracy ODE solver. The resulting state is then projected onto the Pauli basis to extract expectation values, while the Hamiltonian itself is decomposed into its composite Pauli operators. These results are then saved, to be used for displaying the results on the Bloch sphere.

